\documentclass[twocolumn]{aastex62}

\graphicspath{{./}{figures/}}

\shorttitle{Pre-Main Sequence Star Properties}
\shortauthors{Nofi et al.}
\usepackage{verbatim}
\usepackage{upgreek}
\usepackage{afterpage}

\begin{document}

\title{Projected Rotational Velocities and Fundamental Properties of Low-Mass Pre-Main Sequence Stars in the Taurus-Auriga Star Forming Region}

\correspondingauthor{Larissa Nofi}
\email{larissa.nofi@aero.org}

\author{Larissa A. Nofi}
\affiliation{The Aerospace Corporation
2310 E. El Segundo Blvd.,
El Segundo, CA 90245, USA}
\affiliation{Institute for Astronomy, University of Hawai'i at Manoa 
2680 Woodlawn Dr.,
Honolulu, HI 96822, USA}
\affiliation{Lowell Observatory 
1400 W. Mars Hill Rd., 
Flagstaff, AZ 86001, USA}

\author{Christopher M. Johns--Krull}
\affiliation{Department of Physics and Astronomy, Rice University
6100 Main St.,
Houston, TX 77005, USA}
%\collaboration{(AAS Journals Data Scientists collaboration)}

\author{Ricardo L\'{o}pez--Valdivia}
\affiliation{Department of Astronomy, The University of Texas at Austin
2515 Speedway, Stop C1400,
Austin, TX 78712, USA}

\author{Lauren Biddle}
\affiliation{Lowell Observatory 
1400 W. Mars Hill Rd., 
Flagstaff, AZ 86001, USA}
\affiliation{Department of Physics $\&$ Astronomy, Northern Arizona University 
527 S. Beaver St., 
Flagstaff, AZ 86011, USA}

\author{Adolfo S. Carvalho}
\affiliation{Department of Physics and Astronomy, Rice University 
6100 Main St., 
Houston, TX 77005, USA}

\author{Daniel Huber}
\affiliation{Institute for Astronomy, University of Hawai'i at Manoa 
2680 Woodlawn Dr.,
Honolulu, HI 96822, USA}

\author{Daniel Jaffe}
\affiliation{Department of Astronomy, The University of Texas at Austin 
2515 Speedway, Stop C1400, 
Austin, TX 78712, USA}

\author{Joe Llama}
\affiliation{Lowell Observatory 
1400 W. Mars Hill Rd., 
Flagstaff, AZ 86001, USA}

\author{Gregory Mace}
\affiliation{Department of Astronomy, The University of Texas at Austin 
2515 Speedway, Stop C1400, 
Austin, TX 78712, USA}

\author{Lisa Prato}
\affiliation{Lowell Observatory
1400 W. Mars Hill Rd.,
Flagstaff, AZ 86001, USA}

\author{Brian Skiff}
\affiliation{Lowell Observatory 
1400 W. Mars Hill Rd., 
Flagstaff, AZ 86001, USA}

\author{Kimberly R. Sokal}
\affiliation{Department of Astronomy, The University of Texas at Austin 
2515 Speedway, Stop C1400, 
Austin, TX 78712, USA}

\author{Kendall Sullivan}
\altaffiliation{NSF Graduate Research Fellow}
\affiliation{Department of Astronomy, The University of Texas at Austin 
2515 Speedway, Stop C1400, 
Austin, TX 78712, USA}

\author{Jamie Tayar}
\altaffiliation{Hubble Fellow}
\affiliation{Institute for Astronomy, University of Hawai'i at Manoa 
2680 Woodlawn Dr.,
Honolulu, HI 96822, USA}

%\author{Butler Burton}
%\affiliation{National Radio Astronomy Observatory}
%\affiliation{AAS Journals Associate Editor-in-Chief}
%\nocollaboration

%\altaffiliation{Creator of AASTeX v6.2}
%\collaboration{(LaTeX collaboration)}

%\affiliation{AAS Director of Publishing}
%\affiliation{American Astronomical Society \\
%2000 Florida Ave., NW, Suite 300 \\
%Washington, DC 20009-1231, USA}

%\author{Jeff Lewandowski}
%\affiliation{IOP Senior Publisher for the AAS Journals}
%\affiliation{IOP Publishing, Washington, DC 20005}



\begin{abstract}

The projected stellar rotational velocity ($v \sin i$) is critical for our understanding of processes related to the evolution of angular momentum in pre-main sequence stars. We present $v \sin i$ measurements of high-resolution infrared and optical spectroscopy for 70 pre-main sequence stars in the Taurus-Auriga star-forming region, in addition to effective temperatures measured from line-depth ratios, and stellar rotation periods determined from optical photometry. From the literature, we identified the stars in our sample that show evidence of residing in circumstellar disks or multiple systems. The comparison of infrared $v \sin i$ measurements calculated using two techniques shows a residual scatter of $\sim$ 1.8 km s$^{-1}$, defining a typical error floor for the $v \sin i$ of pre-main sequence stars from infrared spectra. A comparison of the $v \sin i$ distributions of stars with and without companions shows that binaries/multiples typically have a higher measured $v \sin i$, which may be caused by contamination by companion lines, shorter disk lifetimes in binary systems, or tidal interactions in hierarchical triples. A comparison of optical and infrared $v \sin i$ values shows no significant difference regardless of whether the star has a disk or not, indicating that CO contamination from the disk does not impact $v \sin i$ measurements above the typical $\sim$ 1.8 km s$^{-1}$ error floor of our measurements. Finally, we observe a lack of a correlation between the $v \sin i$, presence of a disk, and H-R diagram position, which indicates a complex interplay between stellar rotation and evolution of pre-main sequence stars. 

\end{abstract}


\section{Introduction} \label{sec:intro}

The projected stellar rotational velocity ($v \sin i$, where $i$ is the inclination) is an important observable that provides information about the angular momentum evolution of a star. Stellar rotation has been linked to dynamo-driven magnetic activity (e.g., \citealt{1987ARA&A..25..271H}; \citealt{1993MNRAS.265..359G}; \citealt{2013EAS....62..143B}), mass loss through stellar winds (e.g., \citealt{1992MNRAS.256..269T}; \citealt{2005ApJ...632L.135M}; \citealt{2017A&A...598A..24J}), interactions with circumstellar disks (e.g., \citealt{1992ApJ...398L..61A}; \citealt{2003astro.ph..3199M}), disk lifetimes (e.g., \citealt{1997MmSAI..68..881B}), stellar internal structure, differential rotation, internal transport of angular momentum (e.g., \citealt{2004ApJ...601..979W}; \citealt{2015MNRAS.446L..51B}), and stellar birth environments (e.g., \citealt{1996AJ....111..283C}; \citealt{2018arXiv181006106S}). Rotation produces a systematic lowering of the effective temperature of stars which can be as large as $\sim$ 300 K for very rapid rotation (\citealt{2000ApJ...534..335S}). Measurements of $v \sin i$, as a supporting criterion of youth, have been used to aid in membership studies, confirming kinematic members within young open clusters (e.g., \citealt{2010MNRAS.409..552G}). The $v \sin i$ can inform phenomena related to the extended main sequence turn-off, which constrains cluster ages (e.g., \citealt{2018MNRAS.480.3739B}; \citealt{2018AJ....156..165C}; \citealt{2019ApJ...876..113S}). It is also an important parameter when considering interactions between two bodies, such as stellar binary interactions involving disk truncation and the resulting stellar spin-up (e.g., \citealt{2018AJ....156..275S}), and stellar spin history related to planet formation, evolution, and star-planet interactions (e.g., \citealt{2010ApJ...723L..64C}; \citealt{2012A&A...544A.124B}; \citealt{2015MNRAS.453.3720B}; \citealt{2019A&A...621A.124B}). Furthermore, stellar rotation-induced activity and winds cause mass loss of planetary atmospheres, and therefore play a significant role in planet habitability in main sequence systems (e.g., \citealt{2013prpl.conf2G021K}; \citealt{2015ApJ...815L..12J}; \citealt{2016csss.confE.147V}; \citealt{2017IAUS..328..168J}). 

The rotational velocities of pre-main sequence stars have been particularly significant in contributing to our understanding of the evolution of angular momentum in young stars. Young clusters of different ages have been studied to understand the rotational history of low-mass stars. Early observations of pre-main sequence rotational velocities revealed that this population typically rotates more slowly than expected based on stellar evolution theory alone, at a small fraction of the star's break-up velocity (\citealt{1981ApJ...245..960V}; \citealt{1986A&A...165..110B}; \citealt{1989AJ.....97..873H}; \citealt{1990AJ.....99..946B}; \citealt{2000MNRAS.319..457C}). To account for the observed low $v \sin i$ ($\sim$ 15 km s$^{-1}$) of this population, a strong braking mechanism was needed (\citealt{1993A&AS..101..485B}). One proposed mechanism is magnetic disk-locking, in which the star experiences magnetic braking through interactions with the circumstellar disk, causing the young star to lose angular momentum (\citealt{1990RvMA....3..234C}; \citealt{1991ApJ...370L..39K}; \citealt{1992ApJ...398L..61A}; \citealt{2002ApJ...564..877T}; \citealt{2003astro.ph..3199M}). This mechanism is supported by several studies showing that the median $v \sin i$ is lower for those populations that show evidence of hosting disks (\citealt{1993A&A...272..176B}; \citealt{1993AJ....106..372E}; \citealt{1996AJ....111..283C}; \citealt{2002A&A...396..513H}; \citealt{2005AJ....129..363S}; \citealt{2012ApJ...745...56D}; \citealt{2012ApJS..202....7X}). Therefore, the resulting distributions of rotational velocities and disk dispersal timescales can inform disk-coupling lifetimes (\citealt{1993AJ....106..372E}; \citealt{1997MmSAI..68..881B}; \citealt{2000ApJ...534..335S}; \citealt{2002ApJ...564..877T}). Stars with long-lived disks become relatively slow rotators and stars with short-lived disks become relatively fast rotators (\citealt{1997ApJ...475..604S}). An additional braking mechanism of young stars is a magnetically-coupled stellar wind, which results in a significant loss of angular momentum (\citealt{1989AJ.....97..873H}; \citealt{1992MNRAS.256..269T}; \citealt{2012A&A...544A.124B}). Once disk dispersal takes place, the pre-main sequence star spins up as it contracts toward the zero-age main sequence (\citealt{1994ApJ...429..781S}; \citealt{2005MNRAS.356..167M}; \citealt{2013EAS....62..143B}). 

In this paper, we present $v \sin i$ measurements derived from infrared spectroscopy for a sample of 70 pre-main sequence stars. Additionally, we present optically-measured $v \sin i$ estimates, which we compare to the infrared $v \sin i$ values, effective temperatures ($T_{\rm eff}$, also from infrared spectroscopy), stellar rotation periods (P$_{\rm rot}$) from optical photometry, and consequently, lower limits on stellar radii ($R \sin i$). Our sample consists of classical T Tauri stars (CTTSs) and weak-line T Tauri stars (WTTSs), as identified in previously published studies that looked for evidence of an accreting circumstellar disk. The sample includes both single stars or stars in multiple systems, also classified from the literature. We investigate possible trends and correlations between stellar parameters; in particular, we compare the $v \sin i$ distributions of stars with and without disks to test stellar evolution theory and infer the influence of a circumstellar disk on stellar rotation. 

\section{Sample, Observations, and Data Reduction} \label{sec:sam/obs/red}

\subsection{Sample \label{sec:sample}}

Our full sample of 70 pre-main sequence stars of spectral type K and M is located in the $<$ 5 Myr old (\citealt{2009ApJ...704..531K}) Taurus-Auriga star-forming region at a distance of $145^{+12}_{-16}$ pc away (\citealt{2019A&A...624A...6Y}). There exists an older population of stars projected against the 
Taurus star-forming region, with ages $\gtrsim$10 Myr (\citealt{2018AJ....156..271L}). Four of the targets in this sample (NTTS 040142+2150 NE, NTTS 040142+2150 SW, NTTS 040234+2143, and NTTS 041559+1716) have been identified as belonging to this older population (\citealt{2018AJ....156..271L}). About 60$\%$ of our sample show evidence of having an accreting circumstellar disk based on the measured equivalent width of H$\alpha$ lines, and are thus classified as CTTSs, while the remaining targets that show no evidence of this accretion signature are classified as WTTSs. The additional 10$\%$ are not classified based on H emission, but are infrared sources from the IRAS catalog presumed to have disks. We initially selected our sample for a radial velocity search for young substellar companions, particularly hot Jupiters (\citealt{2008ApJ...678..472H}; \citealt{2008ApJ...687L.103P}; \citealt{2012ApJ...761..164C}; \citealt{2016ApJ...826..206J}). Discoveries from this young exoplanet survey will inform planet formation and evolution timescales and mechanisms. The targets for this survey were chosen based on 3 main criteria (\citealt{1988cels.book.....H}): 1) brightness (V $<$ 15); 2) slow rotation ($v \sin i$ $<$ 20 km s$^{-1}$); 3) spectral type (K and M stars). To increase the sample size, we later included several binaries with separations $>$ 0.05$^{\prime\prime}$, as well as additional single stars, which did not necessarily meet the first set of criteria and are either fainter, faster rotators, or have higher mass. Several of these additional targets were drawn from the sample of primarily WTTSs identified for the NASA Space Interferometry Mission, which aimed to search for young planetary systems (\cite{2002swsi.conf....1B}). Table \ref{tab:properties} lists the targets in the sample, as well as their positions, K magnitudes, and multiplicity status primarily taken from \citealt{2012ApJ...745...19K}. The older population targets are also noted in Table \ref{tab:properties}. 

\begin{deluxetable}{cc} 
\tablecaption{Stellar Properties of Pre-Main Sequence Stars  \label{tab:properties}}
\tablewidth{700pt}
\tabletypesize{\scriptsize}
\tablehead{
\colhead{Column Label} & Description  
} 
\startdata
Name & Target name \tablenotemark{a}\tablenotemark{b} \\
RAh & Hour of Right Ascension (J2000) \\
RAm & Minute of Right Ascension (J2000) \\
RAs & Second of Right Ascension (J2000) \\
DE- & Sign of the Declination  \\
DEd & Degree of Declination (J2000) \\
DEm & Arcminute of Declination (J2000) \\
DEs & Arcsecond of Declination (J2000) \\
Kmag & K-band apparent magnitude \\
N-IR & Number of infrared observations \\
N-op & Number of optical observations \\
Mult & Multiplicity \tablenotemark{c} \\
Teff & Effective temperature \tablenotemark{d} \tablenotemark{e} \\
l$\_$vsini-IR & Limit flag on vsini-IR \\
vsini-IR & Infrared $v \sin i$ \tablenotemark{f} \\
e$\_$vsini-IR & infrared $v \sin i$ uncertainty \\ 
l$\_$vsini-op & Limit flag on vsini-op \\
vsini-op & Optical $v \sin i$ \\
e$\_$vsini-op & Optical $v \sin i$ uncertainty \\
vsini-pub & Published $v \sin i$ \tablenotemark{g} \\
e$\_$vsini-pub & Published $v \sin i$ uncertainty \\
Prot & Stellar rotation period \\
e$\_$Prot & Period uncertainty \\
r$\_$Prot & Period reference \tablenotemark{h} \\
l$\_$Rsini & Limit flag on Rsini \\
Rsini & Lower limit stellar radius \\
e$\_$Rsini & $R \sin i$ uncertainty \\
Class & Classification \tablenotemark{i} \\
r$\_$Class & Classification reference \tablenotemark{j} \\
\enddata
\tablenotetext{a}{In the cases where the target name represents a wide binary only the primary component was characterized unless the secondary is also specified. }
\tablenotetext{b}{The following targets belong to an older population with age $>$10 Myr: NTTS 040142+2150 NE, NTTS 040142+2150 SW, NTTS 040234+2143, and NTTS 041559+1716 (\citealt{2018AJ....156..271L}).}
\tablenotetext{c} {single (S), binary (B), tertiary (T), quadruple (Q)}
\tablenotemark{d}{The relations used to determine $T_{\rm eff}$ are limited to values between 3100-4100. All values outside this range are extrapolated.}
\tablenotetext{e}{All $T_{\rm eff}$ uncertainties are 200 K.}
\tablenotetext{f}{IT Tau and LkHa 332/G1 show evidence of possible contamination from a stellar companion, which may lead to an overestimated $v \sin i$.}
\tablenotetext{g}{The published $v \sin i$ values come from Luhman (2018). }
\tablenotetext{h}{The stellar rotation periods were obtained from the following: (1) Lowell Observatory photometry, (2) \textit{K2} data, (3) Bouvier et al. (1995), (4) Artemenko et al. (2012), (5) Grankin et al. (2008), (6) Norton et al. (2007), (7) Donati et al. (2019), (8) Percy et al. (2006). }
\tablenotetext{i}{classical (C) or weak-line (W) T Tauri star.} 
\tablenotemark{j}{The target classification references are as follows: (1) Herbig $\&$ Bell (1988), (2) Kraus et al. (2012) (and references therein), (3) Nguyen et al. (2012), (4) Chavarr\'{i}a-K et al. (2000). } \\ \\
\tablecomments{This table is available in its entirety in machine-readable form.}
\end{deluxetable}

\subsection{Observations and Data Reduction \label{sec:obs/red}}

\subsubsection{Infrared Spectroscopy}

The infrared spectral observations used to measure the $T_{\rm eff}$ and the $v \sin i$ were obtained using the Immersion GRating INfrared Spectrometer (IGRINS) at both the 2.7-m Harlan J. Smith Telescope at McDonald Observatory, and the 4.3-m Lowell Discovery Telescope between October 2014 and March 2018. IGRINS is a high-resolution (R $\sim$ 45,000) near-infrared spectrograph with no moving parts that simultaneously records the H- and K-bands (1.4-2.5 $\upmu$m). Additional discussion of the IGRINS design and capabilities can be found in \citet{2014SPIE.9147E..1DP}. 

The targets were observed by nodding between the AB positions of the slit in either AB pairs or ABBA quads, in order to sample and remove the sky background. Standard A0 telluric stars at similar airmasses to the target observations were also observed and were used to remove the telluric absorption features that are prevalent at near-infrared wavelengths, prior to analysis. 

The data were reduced with the standard IGRINS Python pipeline (v2.2.0 alpha 1) (\citealt{jae_joon_lee_2017_845059}). The pipeline implements bad pixel correction, flat-fielding, sky background subtraction, spectral extraction, spectral distortion correction, and wavelength calibration. The wavelength solution is determined by fitting night sky OH emission lines during a first pass, and telluric H$_{2}$O absorption lines in the A0 telluric star during a second pass. The final output is a telluric-corrected spectrum separated into 43 orders. The SNR of our observations was typically $\geq$ 100. 

In total, we obtained 541 spectra of the targets in the sample. Table \ref{tab:properties} shows the number of infrared spectroscopic observations of each star. We removed any observations with low SNR, poor telluric correction, or extreme-veiling continuum caused by hot dust in the circumstellar disk which weakens the spectral lines. Several targets displayed such strong evidence of veiling in the infrared spectra that we were unable to accurately calculate the $v \sin i$ at the SNR of the observations (CZ Tau, DG Tau, DR Tau, HN Tau, and IRAS F04325+2402), however, we still included these targets in our sample of 70 stars to characterize other properties.

\begin{figure}
\begin{center}
\resizebox{\hsize}{!}{\includegraphics{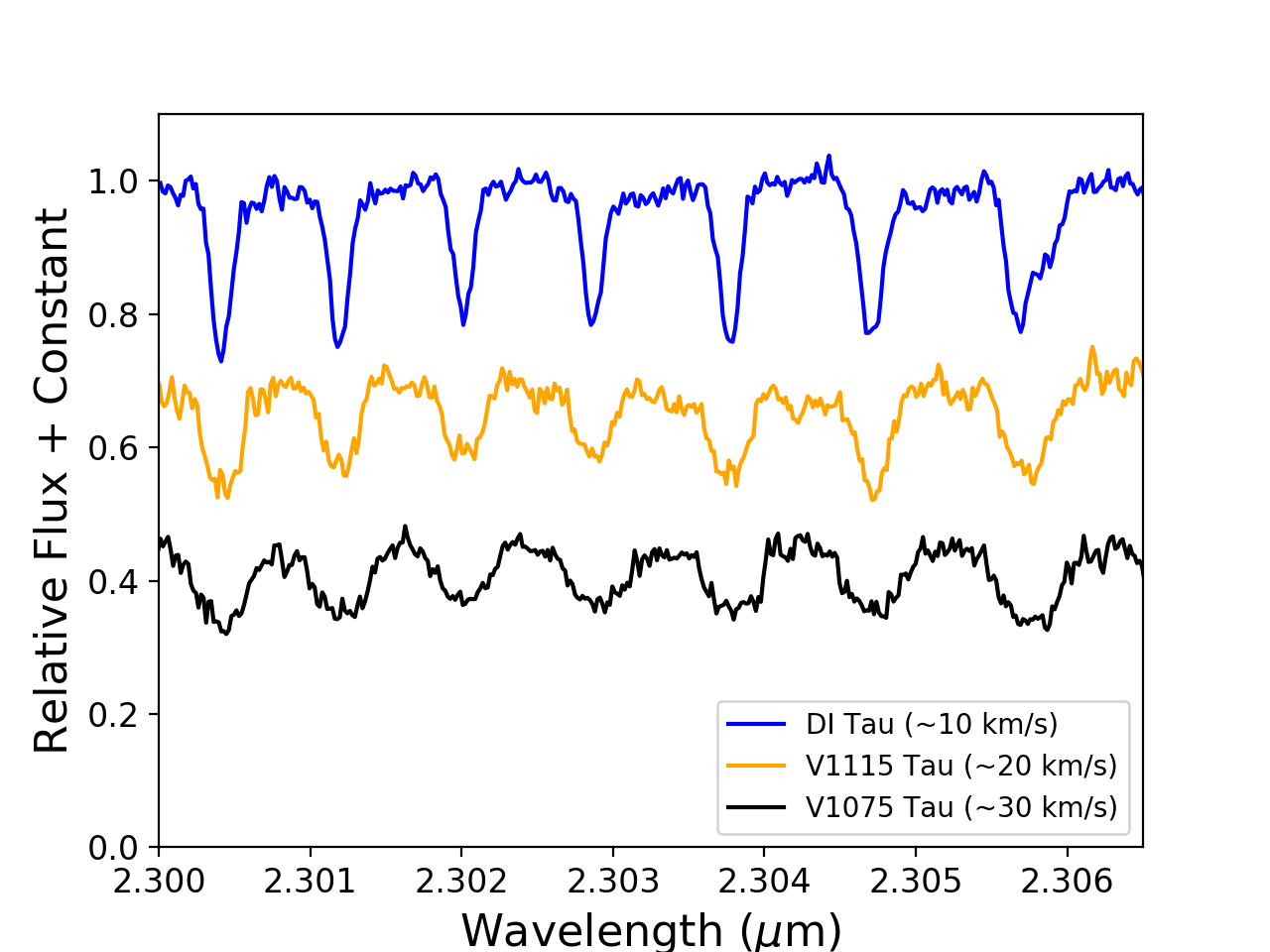}}
\caption{IGRINS K-band continuum-normalized spectra of three targets with a range of $v \sin i$ values. The wavelength region shown is a subset of our analysis region with CO absorption lines. Included are spectra of DI Tau ($v \sin i \sim$ 10 km s$^{-1}$ -- blue), V1115 Tau ($v \sin i \sim$ 20 km s$^{-1}$ -- orange), and V1075 Tau ($v \sin i \sim$ 30 km s$^{-1}$ -- black). The stellar spectral lines become broader as the $v \sin i$ increases. \label{fig:spectra}}
\end{center}
\end{figure}

A sample of our IGRINS spectra is shown in Figure \ref{fig:spectra}. The wavelengths shown represents a subset of the region analyzed in this work. The figure demonstrates how the stellar spectral lines broaden with increasing $v \sin i$ for three of our targets with similar $T_{\rm eff}$. 

\subsubsection{Optical Spectroscopy}

In addition to the infrared spectroscopy, we also obtained optical high resolution ($R \sim 60,000$) spectra at McDonald Observatory as part of the ongoing search for substellar companions to young stars.  Starting in 2004 and continuing to the present time, spectra have been obtained primarily with the McDonald Observatory 2.7-m Harlan J. Smith telescope coupled to the Robert G. Tull cross dispersed coud\'e echelle spectrograph (\citealt{1995PASP..107..251T}).  A smaller number of stars have been observed with the 2.1-m Otto Struve telescope coupled to the Sandiford echelle cassegrain spectrometer (\citealt{1993PASP..105..881M}).  The number of optical spectroscopic observations are indicated in Table \ref{tab:properties}. All optical observations were made with the 2.7-m telescope except for the observations of RW Aur, which were collected using the 2.1-m telescope.  

For observations with the 2.7-m telescope, a 1.2$^{\prime\prime}$ wide slit was used in conjunction with a Tektronix $2080 \times 2048$ CCD to record each spectrum. A total of 54-55 spectral orders were recorded on the CCD for each stellar spectrum. Observations were done roughly once each night during a given observing run (typically 5-10 nights). Before and after each stellar spectrum, a Thorium-Argon comparison lamp spectrum was taken to serve as a reference to determine any instrumental radial velocity shift and to determine the wavelength scale for the observations. Because we are only interested in the $v \sin i$ of each star and not the radial velocity, a single wavelength solution was determined for each night based on the first Thorium-Argon lamp spectrum taken that night.

For observations with the 2.1-m telescope, a 1.1$^{\prime\prime}$ wide slit was used in conjunction with a Reticon $1200 \times 400$ CCD to record each spectrum.  Given the configuration of the folded optics for the cassegrain mounting of the spectrometer and the smaller CCD used, a total of only 18-20 full spectral orders were recorded for each stellar spectrum.  Again, observations were made nightly during a given observing run with Thorium-Argon comparison lamp spectra taken before and after each stellar spectrum.  As above, only a single wavelength scale is determined for each night from the first comparison lamp taken for that night.

For both telescopes, all spectra were reduced with a custom package of IDL echelle reduction routines based largely on the data reduction procedures described by \citet{1994PhDT........16V} and \citet{2000SPIE.4008..720H}. The reduction procedure includes bias subtraction, flat fielding by a normalized flat spectrum, scattered light subtraction, and optimal extraction of the spectrum. The blaze function of the echelle spectrometer is removed to first order by dividing the observed stellar spectra by an extracted spectrum of the flat lamp. Final continuum normalization was accomplished by fitting a 2nd 
order polynomial to the blaze corrected spectra in the regions around the lines of interest for this study. Typical signal-to-noise values for the optical spectra in the regions used in the analysis are $30 - 40$. The wavelength solution for each spectrum was determined by fitting a two-dimensional polynomial to $n\lambda$ as function of pixel and order number, $n$, for the extracted thorium lines observed from the internal lamp assembly.  For the 2.7-m telescope, $\sim 1800$ lines were used in the wavelength solution, while $\sim 600$ lines were used in the solution for the 2.1-m telescope.

\subsubsection{Optical Photometry}

We obtained ground-based photometry for many of our targets to determine stellar rotation periods. The data were supplemented with rotation periods from the literature and high-precision space-based light curves obtained by the \textit{K2} Mission (\citealt{2014PASP..126..398H}). Our ground-based photometry comes from seasonal V-band monitoring of the sample stars using the Lowell 0.7-m robotic telescope. The stars have been generally observed about ten nights per month from September through March each year since 2012, using CCD differential aperture photometry. Additionally, we obtained rotation periods for several of our sample targets from the \textit{K2} long-cadence time-series photometry observed during Campaign 13 between 2017 March 8 and 2017 May 27 UTC. We used the K2SC pipeline (\citealt{2015MNRAS.447.2880A, 2016MNRAS.459.2408A}) to remove instrumental systematics in the data while preserving astrophysical variability of the host star. Additional rotation periods were compiled from the literature, where available. 

\section{Analysis} \label{sec:analysis}

\subsection{Effective Temperatures}

\citet{2019ApJ...879..105L} determined the $T_{\rm eff}$ for 254 K and M main sequence stars from H band IGRINS spectra. They provided two linear relations between $T_{\rm eff}$ and the line-depth ratios (LDR) of Fe I, OH, and Al I absorption lines. These relations offer a simple and accurate measure of $T_{\rm eff}$ in cool stars, and can be used for young stellar objects (YSOs) because the LDRs are insensitive to veiling. \citet{2019ApJ...879..105L} tested one of their relations on 12 members of the nearby ($<$ 60 pc; \citealt{2018yCat.1345....0G}) and young ($\sim$ 7-10 Myr; \citealt{2018ApJ...853..120S}) TW Hydrae Association (\citealt{1997AAS...191.9206K}). They found hotter temperatures (varying by $\sim$ 140 K) for stars with $T_{\rm eff}$ between 3200 and 3800 K compared with previous determinations. This discrepancy may be the result of differences in surface gravity between YSOs and the main sequence stars with which the \citet{2019ApJ...879..105L} approach was calibrated. The $T_{\rm eff}$ - LDR relations are valid between $\sim$ 3100 and 4100 K. Most of our sample stars have $T_{\rm eff}$ within the applicability limits of this approach (see Table \ref{tab:properties}). 

We measured the line-depths of the Fe I ($\lambda$ $\sim$ 1.56216 $\upmu$m) line and the OH ($\lambda$ $\sim$ 1.56270 $\upmu$m) doublet, in the same way as \citet{2019ApJ...879..105L}. We then computed the LDR(Fe/OH), which we used in the linear equation $T = 520 \ \times$ LDR(Fe/OH) $+ \ 3230$ K. For the 28 targets with $T_{\rm eff}$ outside the limits of the relation, we extrapolated the value of $T_{\rm eff}$. A typical measurement error in the LDR analysis is $\sim$ 140 K, and a typical systematic error is $\sim$ 120 K. Therefore,  we determined a total uncertainty of $\sim$ 180 K and have assigned a conservative error of 200 K for the whole sample.

As this method was applied to a new regime of stellar age, we made a comparison between the $T_{\rm eff}$ estimates derived from LDRs, and the $T_{\rm eff}$ estimates derived from a relation for Taurus stars that converts spectral type to temperature (\citealt{2014ApJ...786...97H}). We compared the $T_{\rm eff}$ measurements using both methods for an overlapping subset of 57 targets. The differences in $T_{\rm eff}$ were within our 200 K uncertainties for $\sim$ 80$\%$ of the sub-sample (and 90$\%$ of the sub-sample when excluding extrapolated values beyond 4100 K). The mean difference in $T_{\rm eff}$ was $\sim$ 100 K, with a standard deviation of $\sim$ 200 K. This variation in $T_{\rm eff}$ is not expected to significantly impact the $v \sin i$ measurements (see Section \ref{sec:analysis_ir} for details).

\subsection{Infrared $v \sin i$\label{sec:analysis_ir}}

Our infrared $v \sin i$ analysis used the IGRINS K-band order with wavelength range 2.299-2.319 $\upmu$m. This region contains CO absorption lines, which are less sensitive to magnetic fields or pressure broadening (\citealt{2018AJ....155..225K}). This region also contains telluric absorption lines, which we removed using A0 star division. Magnetically-sensitive Ti I lines in this region were masked to avoid measuring broadening of the lines from magnetic fields. The final spectrum was continuum-normalized and trimmed to avoid inaccuracies caused by distortion at the ends of the spectral order. 

\begin{figure}
\begin{center}
\resizebox{\hsize}{!}{\includegraphics{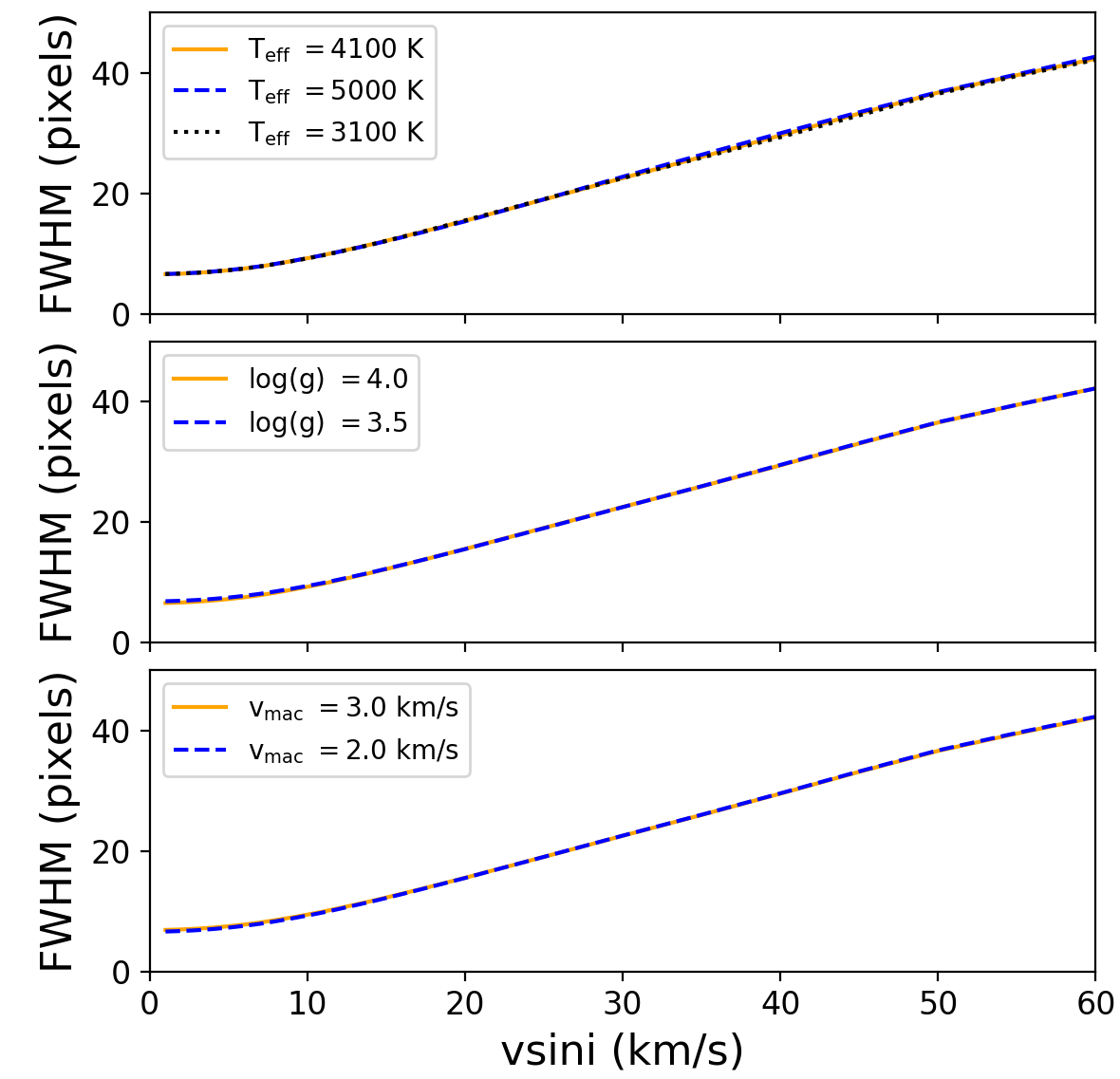}}
\caption{Relations between the FWHM of the CCF and $v \sin i$ for different stellar parameters: effective temperature (top), surface gravity (middle), and macroturbulence velocity (bottom). Using these relations, we determined the $v \sin i$ from a measurement of the FWHM of the CCF, which indicates the level of rotational broadening of the spectral lines. The relations do not vary significantly based on differences in the model parameters. \label{fig:relation}}
\end{center}
\end{figure}

The infrared $v \sin i$ values were measured using the technique outlined in \citet{1986ApJ...309..275H} and \citet{1989AJ.....97..539S}. The basic premise of this technique is that when a stellar spectrum with rotationally broadened lines is cross-correlated against an unbroadened, narrow-lined spectrum, the resulting width of the cross-correlation function (CCF) can be used to measure the rotational broadening of the first spectrum. To measure the $v \sin i$ of our targets, we first calibrated a relation between the full width at half maximum (FWHM) of the CCF and the $v \sin i$. We artificially broadened synthetic spectra over a range of $v \sin i$ from 1 to 20 km s$^{-1}$ in steps of 1 km s$^{-1}$, then up to a $v \sin i$ of 60 km s$^{-1}$ in steps of 5 km s$^{-1}$. This generated a calibration function that relates the FWHM of a CCF to the $v \sin i$. The process was repeated with stellar models of different temperatures to establish calibration relations that matched the range of temperatures of our targets. An unbroadened synthetic spectrum with a similar $T_{\rm eff}$ to the target star was then chosen and cross-correlated with the target spectrum to measure the FWHM of the CCF, which can then be related to the corresponding $v \sin i$ from the calibration relation. Figure \ref{fig:relation} shows calibration relations for several synthetic spectra used in our analysis. While other phenomena can broaden spectral lines in addition to rotation, the FWHM of the CCF is a good indicator of $v \sin i$ for our sample because pre-main sequence stars have relatively low gravities in their atmospheres, which results in narrow photospheric absorption features in which rotational broadening dominates over pressure broadening (\citealt{2012ApJ...745...56D}).

Typically, the FWHM of the CCF is measured from a baseline of zero. However, the stellar absorption lines in our region of interest are nearly evenly-spaced (see Figure \ref{fig:spectra}), which creates a sinusoidal pattern in the CCF, causing the baseline of the CCF response to be negative. To accurately measure the FWHM we instead measured from a baseline determined by the minimum of the CCF response. To account for variations due to noise, we calculated the average of the two minima on either side of the CCF peak (within $\pm$100 pixels, which corresponds to about $\pm$200 km s$^{-1}$) as the baseline for the FWHM measurement. This modification was made both when creating the calibration relations, and measuring the FWHM of the CCF from the dataset to determine the  $v \sin i$. 

While several studies have determined $v \sin i$ with this technique, most rely on observed spectral templates (\citealt{2001AJ....122.3258R}; \citealt{2004ApJ...601..979W}; \citealt{2006AJ....132.1555N}; \citealt{2009ApJ...695.1648N}; \citealt{2012ApJ...745...56D}). Our work relies on synthetic spectra at infrared wavelengths to create a calibration relation and measure the $v \sin i$ of our targets. There is a good match between observational and theoretical spectra in the K-band (\citealt{2012MNRAS.422.2195L}). Rotational velocity standard templates may have significantly different physical properties than the target stars (\citealt{2012MNRAS.422.2195L}) and are non-zero rotators, which can lead to underestimated rotational velocities. However, this has been shown to be an insignificant effect for stars with $v \sin i$ $\geq$ 35 km s$^{-1}$, and a small effect ($<$ 5 km s$^{-1}$) for stars with $v \sin i$ $<$ 25 km s$^{-1}$ (\citealt{1981ApJ...245..960V}). Despite theoretical support, in practice templates are a standard method used for measuring $v \sin i$ that likely introduces negligible errors (see Section 3.3). 

Synthetic infrared spectra covering the region of interest were computed using the SYNTHMAG spectrum synthesis code (\citealt{1999ASSL..243..515P}).  Line data needed to compute the synthetic spectra were obtained from the Vienna Atomic Line Database (\citealt{1995A&AS..112..525P}; \citealt{2015BaltA..24..453R}).  The region of interest is dominated by lines of the CO molecule, and the line data in VALD for this molecule comes from \citet{1994ApJS...95..535G}.  We use the NextGen stellar atmosphere models (\citealt{1995bmsb.conf...32A}) for the synthetic spectrum calculation.  In all cases a microturbulent broadening of 1 km s$^{-1}$ is assumed (\citealt{2011A&A...526A.103D}), and the macroturbulent broadening (discussed below) uses the radial-tangential formulation (\citealt{2008oasp.book.....G}) with the radial and tangential turbulent velocities equal. The synthetic spectra were convolved with a Gaussian to represent the standard resolution of the IGRINS spectra (R $\sim$ 45,000) before they were used as a template to measure $v \sin i$. We generated the synthetic models at varying $T_{\rm eff}$, surface gravity ($\log g$), and macroturbulence velocity ($v_{\rm mac}$). Metallicity remained fixed and we assumed solar metallicity for our sample (\citealt{2011A&A...526A.103D}). We conducted tests to investigate which of these stellar parameters are significant to the outcome of the measured $v \sin i$. We explored a range of temperatures between 3100 K and 5100 K (corresponding to the range of $T_{\rm eff}$ estimates of our targets) in steps of 200 K (the uncertainty in the $T_{\rm eff}$ measurements), a $\log g$ of 3.5 and 4.0, and a $v_{\rm mac}$ of 2 km s$^{-1}$ and 3 km s$^{-1}$. Figure \ref{fig:relation} shows how the relation for determining $v \sin i$ based on the FWHM of the CCF changes with $T_{\rm eff}$ (a), $\log g$ (b) and by $v_{\rm mac}$ (c). Varying the $T_{\rm eff}$, $\log g$, and $v_{\rm mac}$ did not significantly contribute to the total uncertainty within the range of parameter values of our targets. For our measurements, we assumed a $v_{\rm mac}$ of 2 km s$^{-1}$, and a $\log g$ of 3.5, since our sample contains cool, low-mass stars (\citealt{2008oasp.book.....G}). 

To obtain the infrared $v \sin i$ measurements, we first matched each target with a synthetic spectrum based on the measured $T_{\rm eff}$. We measured the $v \sin i$ for each individual epoch and calculated the average to determine a final $v \sin i$ value for each target. We determined a random internal uncertainty by measuring the standard deviation of the mean, which on average resulted in an error of $\sim$ 0.7 km s$^{-1}$. 

\begin{figure}
\begin{center}
\resizebox{\hsize}{!}{\includegraphics{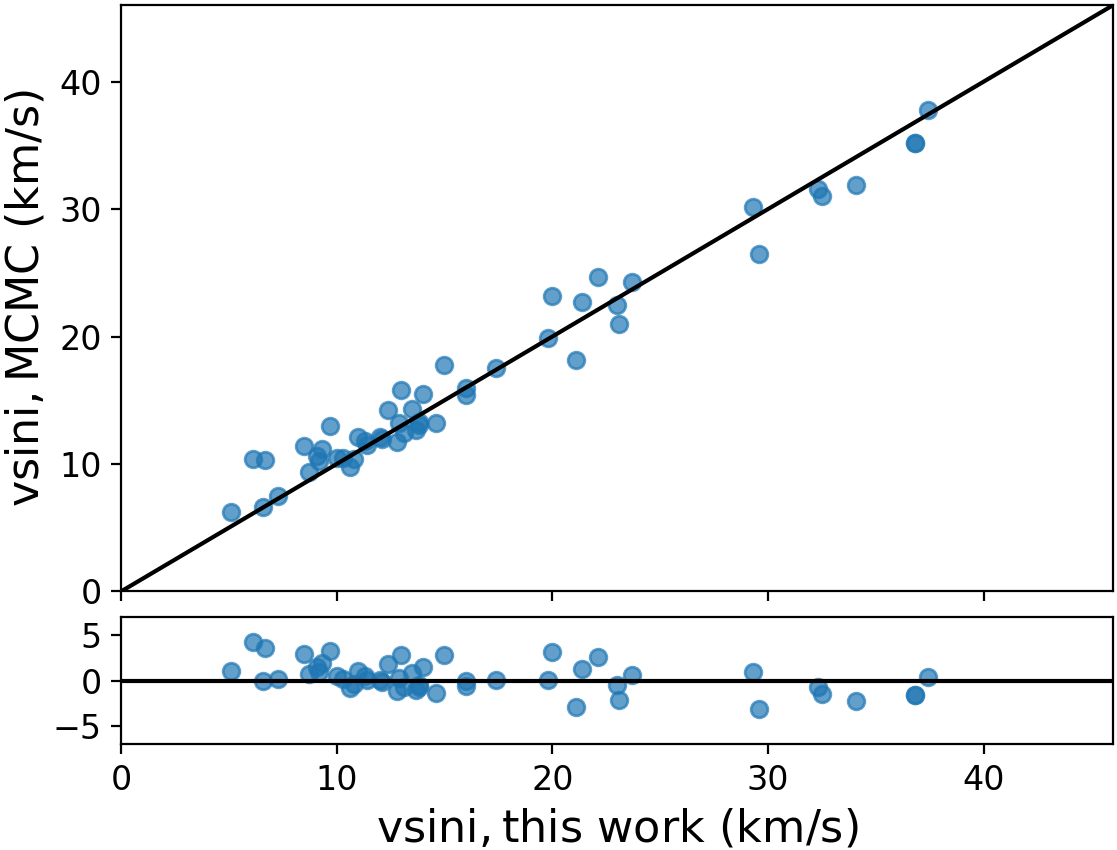}}
\caption{A comparison of $v \sin i$ measurements and residuals using IGRINS spectra and two different methods. L\'{o}pez-Valdivia et al. (in prep.) used an MCMC technique and MOOGStokes model spectra, while this work relies on a cross correlation technique and NextGen model spectra. We observe a median offset of 1.0 km s$^{-1}$ and residual standard deviation of $\sim$ 1.8 km s$^{-1}$, which we add in quadrature to our formal uncertainties. \label{fig:RLV_LN}}
\end{center}
\end{figure}

Additionally, we considered systematic errors due to different models and methodologies. The most likely source of systematic error is model choice mismatch (\citealt{1986ApJ...309..275H}). While the uncertainty based on model choice can be significant (\citet{2012MNRAS.422.2195L} found that deviations of $T_{\rm eff}$, $\log g$, and $v_{\rm mac}$ lead to uncertainties in the determination of $v \sin i$ at the level of 1-2 km s$^{-1}$). Given the consistency between the calibration relations based on alternate model choices (see Figure \ref{fig:relation}), we determined that the systematic error is negligible in this analysis. We also characterized a systematic error due to different methodologies by comparing the $v \sin i$ values to preliminary measurements from L\'{o}pez-Valdivia et al. (in prep.), who measured the $v \sin i$ for an overlapping sub-sample of 51 targets. Instead of a cross correlation technique, they used a Markov Chain Monte Carlo (MCMC) algorithm to compare observations with MOOGStokes synthetic spectra (\citealt{2013AJ....146...51D}). They also included additional spectral regions in their analysis. Both analyses relied on IGRINS data: this work measured the $v \sin i$ for single spectra and averaged the result, while L\'{o}pez-Valdivia et al. (in prep.) measured $v \sin i$ from a single averaged spectrum. Figure \ref{fig:RLV_LN} shows a comparison of the preliminary MCMC $v \sin i$ estimates and those from this work. There is good agreement between the two results, and the standard deviation of the residuals indicates a methodology uncertainty of $\sim$ 1.8 km s$^{-1}$. This methodology error was added in quadrature to the internal random uncertainty to determine a total uncertainty.

Rotational velocity measurements are limited by the SNR and spectral resolution of the instrument. To determine this limit, we measured the $v \sin i$ of a radial velocity standard star, GJ 281 ($v \sin i$=1.0$\pm$0.9 km s$^{-1}$; \citealt{Smith...dissertation}) which indicated a lower limit of $\sim$ 3-4 km s$^{-1}$. This value is consistent with detection thresholds reported for similar resolution spectra (\citealt{2010AJ....139..504B, 2012AJ....143...93R, 2015AJ....149..106D, 2018AJ....155..225K}). Any $v \sin i$ estimate below this value is considered unreliable, however, we did not measure any $v \sin i$ values below this limit in our sample. 

The broadening of the CCF also becomes difficult to accurately measure for fast rotators. We estimated an upper limit on the $v \sin i$ for stars with high rotational velocities at which we can no longer accurately measure the $v \sin i$. We established this limit by cross-correlating an artificially broadened model, generated with noise corresponding to a SNR of 100, with the original unbroadened synthetic model for different values of $v \sin i$. At $\sim$ 50 km s$^{-1}$ we recover a $v \sin i$ that is inaccurate at the level of the median random uncertainty measured for our dataset. Therefore, we established an upper limit on fast rotators of 50 km s$^{-1}$. 

\subsection{Optical $v \sin i$}\label{sec:analysis_op}

To measure $v \sin i$ values from the optical spectra we used essentially the same technique described above with two exceptions.  The first regards the template, for which we used the observed spectrum of HD 65277 instead of a synthetic spectrum.  HD 65277 is classified as a K4 dwarf star (\citealt{1999MSS...C05....0H}), which implies an effective temperature of 4620 K (\citet{2013ApJS..208....9P}).  \citet{2005yCat..21590141V} included HD 65277 in their spectroscopic analysis of 1040 F, G, and K dwarfs, finding an effective temperature of 4741 K and a $v \sin i$ = 1.0 km s$^{-1}$.  The recent spectroscopic analysis of \citet{2018A&A...615A..76S} finds a stellar $T_{\rm eff}$ = $4660 \pm 14$ K, along with a $v \sin i$ = 1.81 $\pm$ 0.08 km s$^{-1}$.  As these estimates of the rotational velocity are below the resolution ($\sim 2.5$ km s$^{-1}$) of the optical data, HD 65277 can serve as essentially a non-rotating template with a spectral type very similar to our sample (we discuss the impact on our measurements of its non-zero $v \sin i$ below).  

The second difference is that we employ the least squares deconvolution (LSD) technique (\citealt{1997MNRAS.291..658D}) to boost the signal-to-noise of the individual observations.  The LSD technique combines data from many spectral lines in a way similar to a cross-correlation analysis.  The basic idea is the assumption that the observed spectrum is the convolution between a spectrum composed of delta functions at the wavelength of each line and a single broadening function which takes into account all sources of line broadening including stellar rotation and the instrumental profile of the spectrograph.  The strength of the different delta functions is directly proportional to the expected depth of the respective lines. We used the LSD code described in \citet{2013ApJ...776..113C} to create the LSD profiles.  The line list used (rest wavelengths and predicted depths) is based on the list constructed by \citet{2013ApJ...776..113C} for the analysis of BP Tau, a K7 CTTS, but trimmed down to 375 lines spanning the wavelength range from 528.8 nm to 886.6 nm.  The LSD technique then deconvolves the observed spectrum with the line list to produce a high signal-to-noise Stokes I profile, and it is this LSD profile that we measure the FWHM of to relate to the $v \sin i$ of the star.

To construct the relationship between the FWHM of the LSD profile and the stellar $v \sin i$ we artificially broadened the observed spectrum of HD 65277 with a standard rotational broadening kernel (e.g., \citealt{2008oasp.book.....G}). We then computed the LSD profile of this rotationally broadened spectrum and measure the FWHM of the LSD profile.  This procedure was repeated for several values of $v \sin i$ to produce a calibration curve similar to Figure \ref{fig:relation}.  For each pre-main sequence star, we then computed its LSD profile and linearly interpolated on the calibration relationship to determine its $v \sin i$.  Most of our stars have multiple observations, so we repeated this procedure for every observation and took the mean of the different $v \sin i$ values as our final value.  We also computed the standard deviation of the mean for the multiple values and took this as the random uncertainty associated with our measurement. For a few stars where we only have one observation, we looked at the standard deviation of other stars in our sample with similar values for the $v \sin i$ and assigned that as the random uncertainty for these stars. We estimated the systematic uncertainty in our analysis by using two additional slowly rotating stars as templates, achieving agreement to typically better than 0.7 km s$^{-1}$, which we took as our systematic uncertainty that we added in quadrature to the random uncertainties. We can reliably measure $v \sin i$ from the optical spectra down to 4 km s$^{-1}$ and up to 50 km s$^{-1}$. 

Finally, we tested whether the $v \sin i$ of HD 65277 may slightly bias our results. This effect would be stronger at lower values of $v \sin i$ and would cause the measured $v \sin i$ to be underestimated. To first order, we expect the $v \sin i$ of HD 65277 to add in quadrature to the values we used when creating the calibration relationship between FWHM and $v \sin i$, meaning that it would be appropriate to subtract in quadrature the $v \sin i$ of HD 65277 from the values in Table \ref{tab:properties}. Using the larger literature value we find that the maximum correction would correspond to ~0.3 km s$^{-1}$, which is well below our typical uncertainty. The values we report are not corrected for this potential effect. 

\subsection{Stellar Rotation Periods} 

Ground-based photometry from the Lowell 0.7-m robotic telescope was analyzed using a Fourier-fitting routine (\citealt{1989Icar...77..171H}). This technique uses a phase-dispersion minimization routine to identify peaks in the periodogram relating to periodicity in the photometric data. We searched for periodicities between 0.2 and 100 days in 0.01 day increments. A single, persistent period is chosen based on the results of the minimization routine. The period has an uncertainty of $\sim$ 0.1 days, based on season-to-season scatter in the fitted lightcurves. Additional details of the Lowell photometry analysis can be found in \cite{2016ApJ...826..206J}.

\begin{figure*}
\begin{center}
\resizebox{\hsize}{!}{\includegraphics{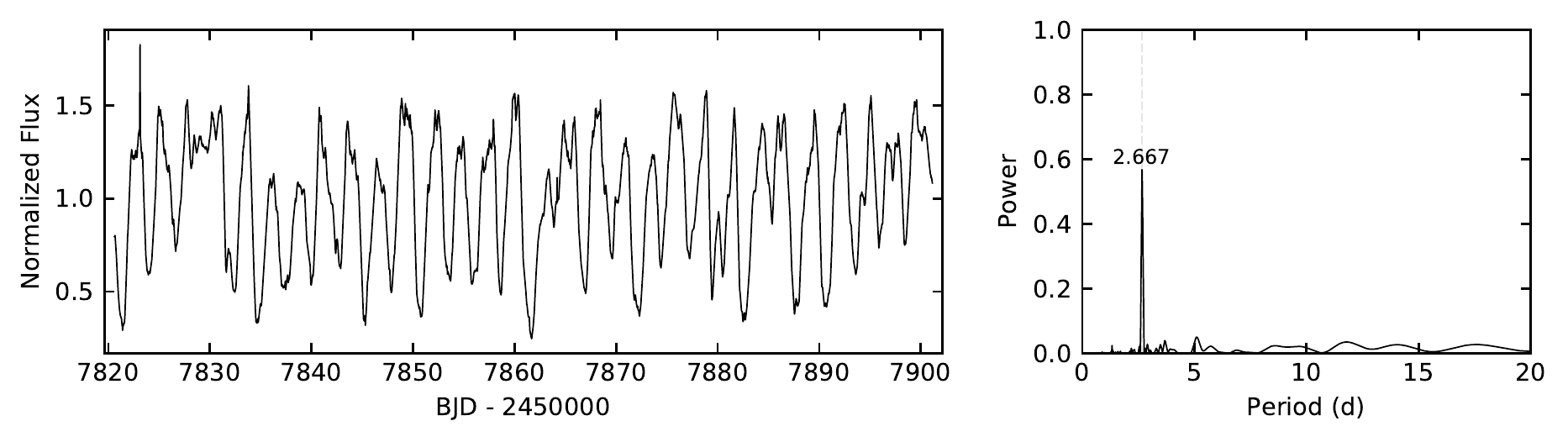}}
\caption{\textit{K2} lightcurve (left) and power spectrum (right) of GM Tau, indicating a stellar rotation period of 2.67 days. \label{fig:K2}}
\end{center}
\end{figure*}

The \textit{K2} data from Campaign 13 are analyzed the same way to measure stellar rotation periods for six of our targets for which we did not obtain rotation periods from ground-based data. We computed a generalized Lomb-Scargle periodogram (\citealt{2009A&A...496..577Z}) of each of the lightcurves, and searched 10,000 points between 0.042 and 80 days (consistent with the Nyquist sampling frequency; \citealt{1992nrca.book.....P}) to ensure that the peaks in the periodogram were well resolved. Distinct periodic signals of rotation for each target were identified, with no strong peaks found at periods beyond 20 days. The rotation periods were determined from the maximum peaks. \citet{2020AJ....159..273R} recently presented rotation periods, also derived from the \textit{K2} Campaign 13 data, which are in agreement with our measurements. V710 Tau A is an exception: we measure a period of $\sim$ 4.0 days instead of the 4.3-days from \citet{2020AJ....159..273R}. We estimated the period uncertainties from the FWHM of the peak, corresponding to the frequency resolution (\citealt{2014sdmm.book.....I}). We used both the analytic solution (\citealt{2009A&A...496..577Z}) and a Monte Carlo bootstrap algorithm to calculate the false-alarm probabilities (FAP). Both methods yielded FAPs of $<$ 10$^{-6}$ for all relevant periods. An example \textit{K2} lightcurve and power spectrum is shown in Figure \ref{fig:K2}. 

Not all photometric observations determined a conclusive rotation period due to quasiperiodic variability. Irregular photometric variations in CTTSs primarily result from accretion from the circumstellar disk onto the star that varies in both time and location (e.g., \citealt{2007prpl.conf..297H}). We do not report rotation periods for several active accretors among the CTTSs, which only show transient periodic behavior (AA Tau, CW Tau, DK Tau, FM Tau, FZ Tau, GM Aur, HN Tau, HO Tau, IQ Tau, V710 Tau A). 

\subsection{Stellar Radius Limits}

Given our measured $v \sin i$ and rotation periods, we were able to place lower limits on the stellar radius using the relation:

\begin{equation}
R \sin i = \frac{v \sin i \ P_{\rm rot}}{(2\pi)}.
\end{equation}

We calculated this value for those stars that have both $v \sin i$ and P$_{\rm rot}$ estimates and propagated the errors from $v \sin i$ and P$_{\rm rot}$, where available (many P$_{\rm rot}$ estimates from the literature did not include uncertainties). We used the infrared $v \sin i$ measurements to determine these values. 

\section{Discussion} \label{sec:discussion}

\subsection{Distributions of Stellar Parameters} 

\begin{figure*}
\centering\offinterlineskip 
\gridline{\fig{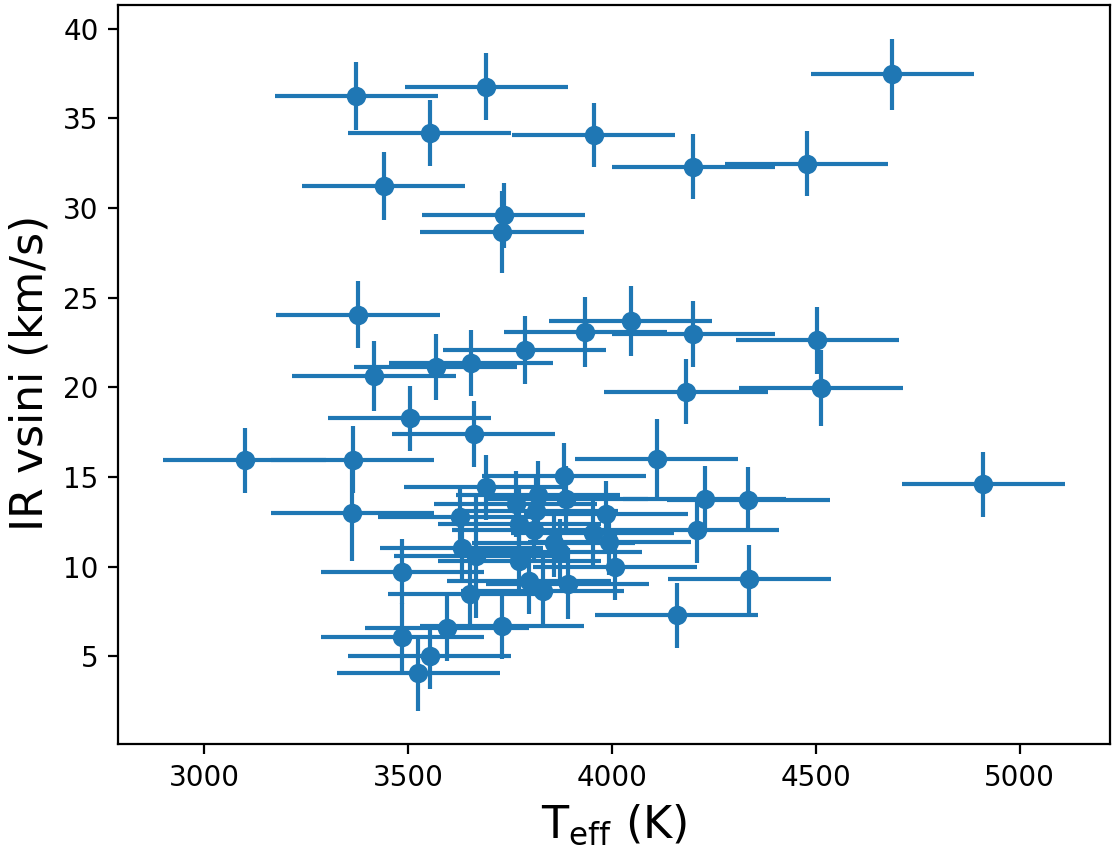}{0.45\textwidth}{(a)}
          \fig{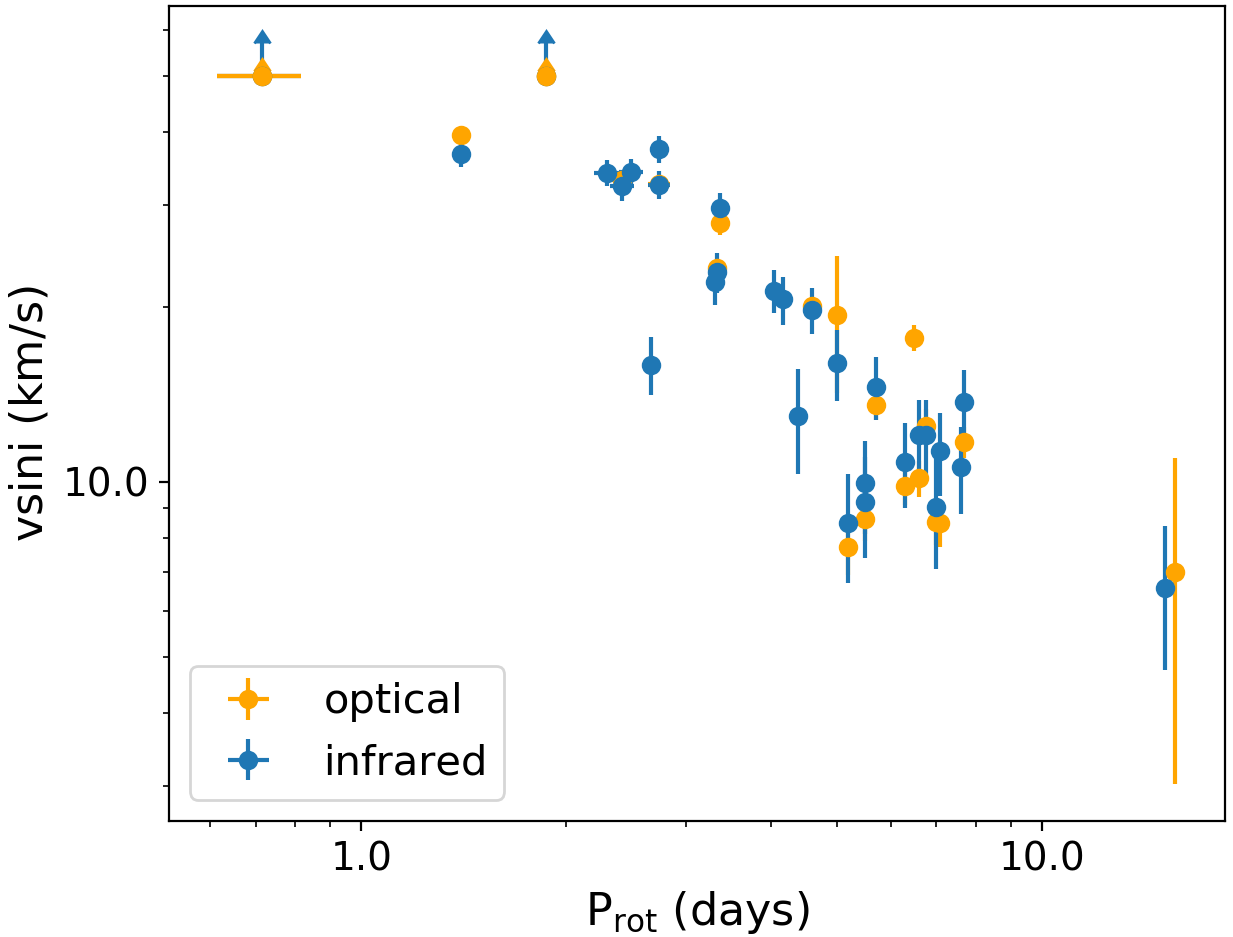}{0.45\textwidth}{(b)}
          }
\gridline{\fig{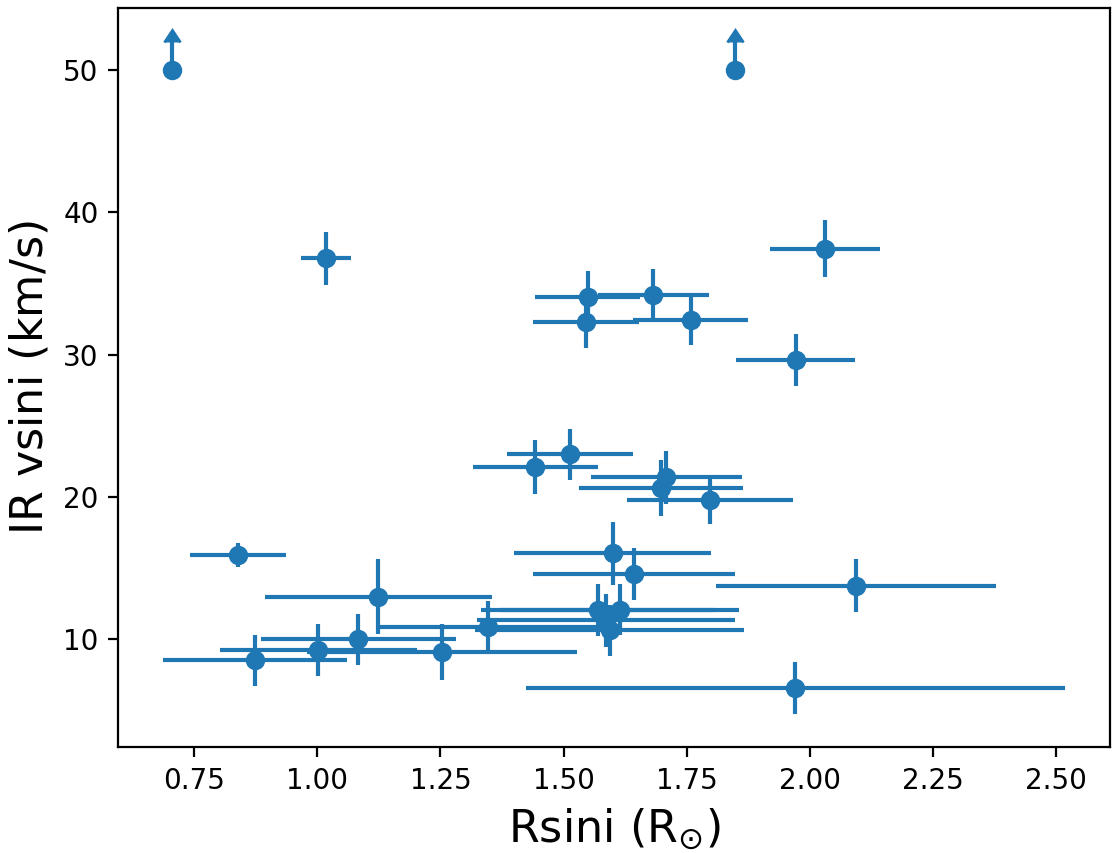}{0.45\textwidth}{(c)}
          \fig{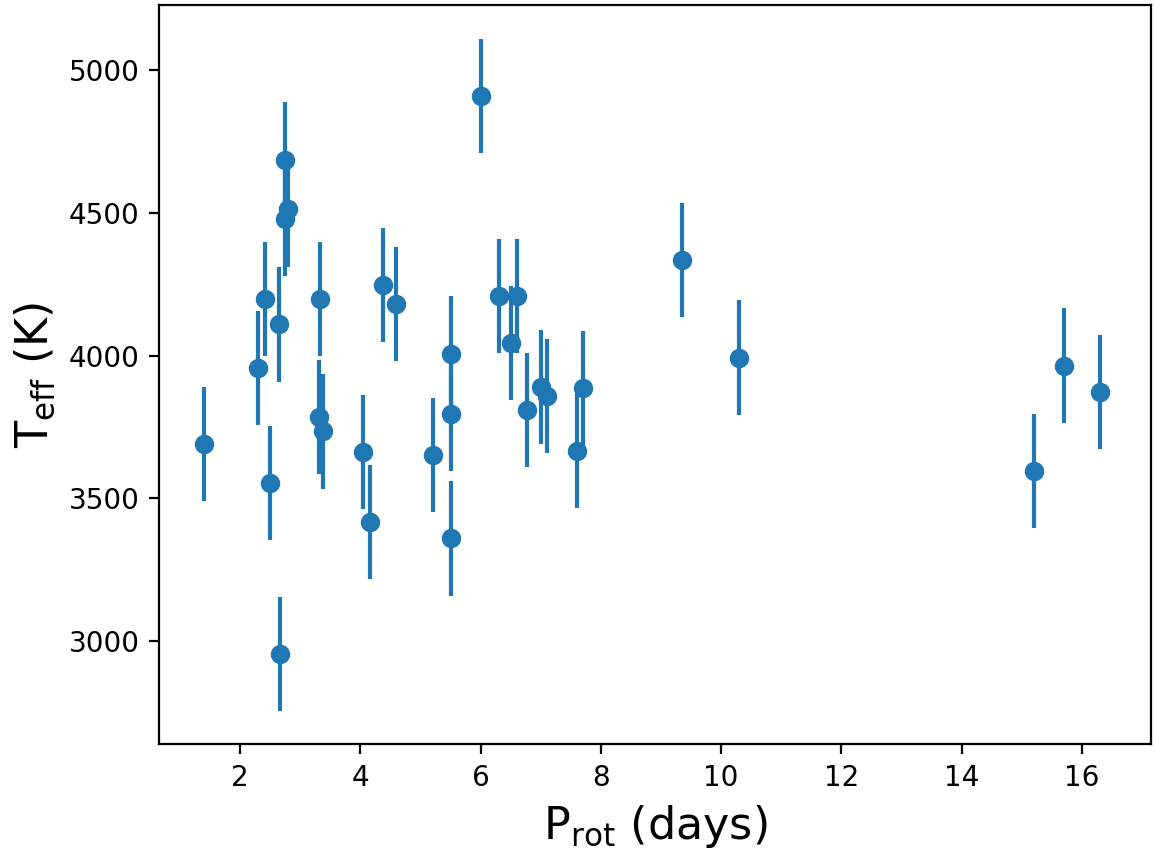}{0.45\textwidth}{(d)}
          }
\gridline{\fig{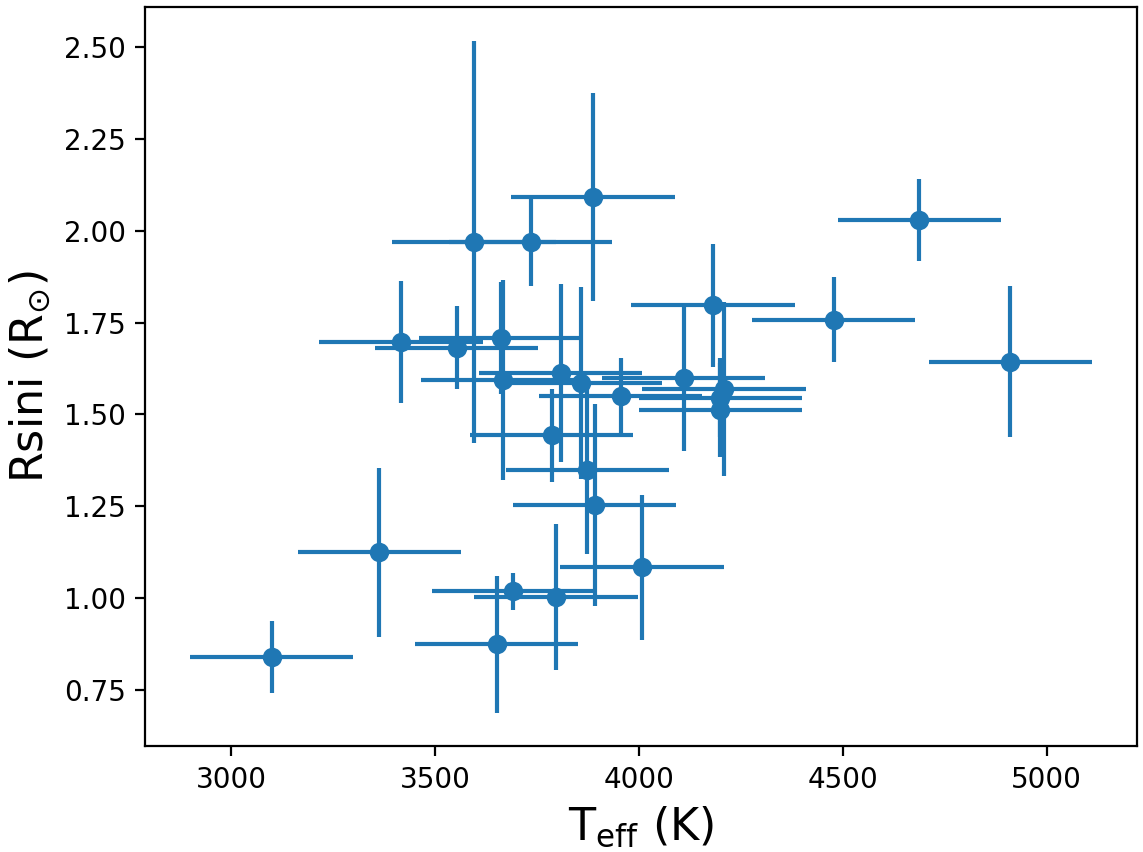}{0.45\textwidth}{(e)}
          \fig{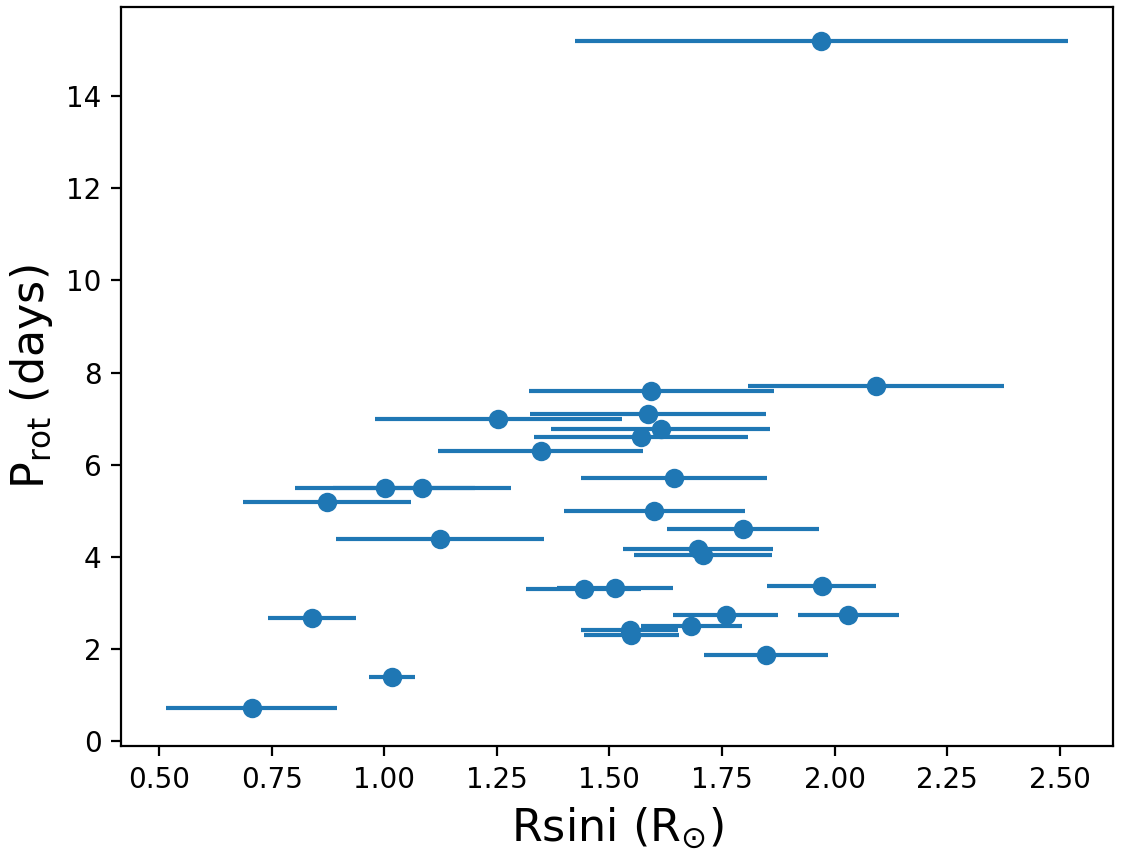}{0.45\textwidth}{(f)}
          }
\centering
\caption{Comparison of the stellar properties measured in this work. Upward arrows indicate limits for fast rotators ($>$ 50 km s$^{-1}$). The P$_{\rm rot}$ error bars are smaller than the points in the figures.} \label{fig:pars_comp}
\end{figure*}

Table \ref{tab:properties} lists $T_{\rm eff}$, infrared and optical $v \sin i$, P$_{\rm rot}$, and $R \sin i$ measurements for our sample. Figure \ref{fig:pars_comp} shows a comparison between the various stellar properties. As expected, the $v \sin i$ shows a strong anti-correlation with the stellar rotation period (Figure \ref{fig:pars_comp}b). There are no obvious correlations among the other stellar parameters. This result hints at a more complicated relationship between evolutionary state and rotation than for normal main sequence stars (see Section \ref{sec:evolution}). The typical $T_{\rm eff}$ and $R \sin i$ is 3800 K and $>$ 1 $R_{\sun}$, respectively, as expected for low-mass pre-main sequence stars contracting down onto the main sequence. 

\begin{deluxetable}{ccCrlc}[b!]
\tablecaption{K-S and A-D Test Results \label{tab:KS}}
\tablecolumns{4}
\tablenum{2}
\tablewidth{0pt}
\tablehead{
\colhead{Comparison} &
\colhead{N} &
\colhead{K-S} & \colhead{A-D} \\
& \colhead{} & \colhead{p-value} & \colhead{p-value}
}
\startdata
single/multiple $v \sin i$ & 40/29 & 0.056 & 0.054 \\
IR/optical $v \sin i$ & 26/26 & 0.440 & 0.405 \\
IR/optical CTTS $v \sin i$ & 19/19 & 0.526 & 0.336 \\
IR/optical WTTS $v \sin i$ & 7/7 & 0.938 & 0.990 \\
IR/published (HS89) $v \sin i$ & 21/21 & 0.135 & 0.207 \\
IR/published (N12) $v \sin i$ & 32/32 & 0.795 & 0.812 \\
optical/published (N12)  $v \sin i$ & 22/22 & 0.175 & 0.086 \\
IR/published (N12) CTTS $v \sin i$ & 20/20 & 0.497 & 0.250 \\
IR/published (N12) WTTS $v \sin i$ & 12/12 & 0.433 & 0.250 \\
CTTSs/WTTS $v \sin i$ & 36/20 & 0.002 & 0.003 \\
CTTSs/WTTS $v \sin i$ (N12) & 52/78 & 0.177 & 0.179 \\
combined CTTSs/WTTS $v \sin i$ & 88/98 & 0.028 & 0.015 \\
\enddata
\tablecomments{Summary of the K-S and A-D test results for various $v \sin i$ comparisons in our sample. The number of measurements and the p-values for both statistical tests are also listed. HS89 refers to \citet{1989AJ.....97..873H} and N12 refers to \citet{2012ApJ...745..119N}.} 
\end{deluxetable}

\subsection{Effects of Multiplicity on $v \sin i$} 

Measurements of $v \sin i$ in binary stars can differ from single stars either through a measurement bias (line blending) or physical effects such as tidal-spin up (e.g., \cite{1992ApJ...398L..61A}. The physical separations of the multiples in our sample span a wide range between $<$ 10 au and several hundred au (\citealt{2012ApJ...745...19K}). Only $\sim$ 1/3 of these systems are expected to be resolved by IGRINS. \cite{2009ApJ...704..531K} report K-band contrast ratios between 0.7-0.9 for $\sim$ 1/4 of these unresolved systems. This indicates that there may be a measurement bias toward higher $v \sin i$ in our sample of multiple stars.

To determine if these broadening effects are present in our data, we compared the average $v \sin i$ of the stellar binaries (or multiples) in our sample with that of the single stars. We found that the average $v \sin i$ of the 28 pre-main sequence stars in multiple star systems is 19.3 $\pm$ 1.8 km s$^{-1}$, while the 34 single stars yield an average $v \sin i$ of 15.4 $\pm$ 1.7 km s$^{-1}$, indicating that these samples are statistically different. 

\begin{figure}
\begin{center}
\resizebox{\hsize}{!}{\includegraphics{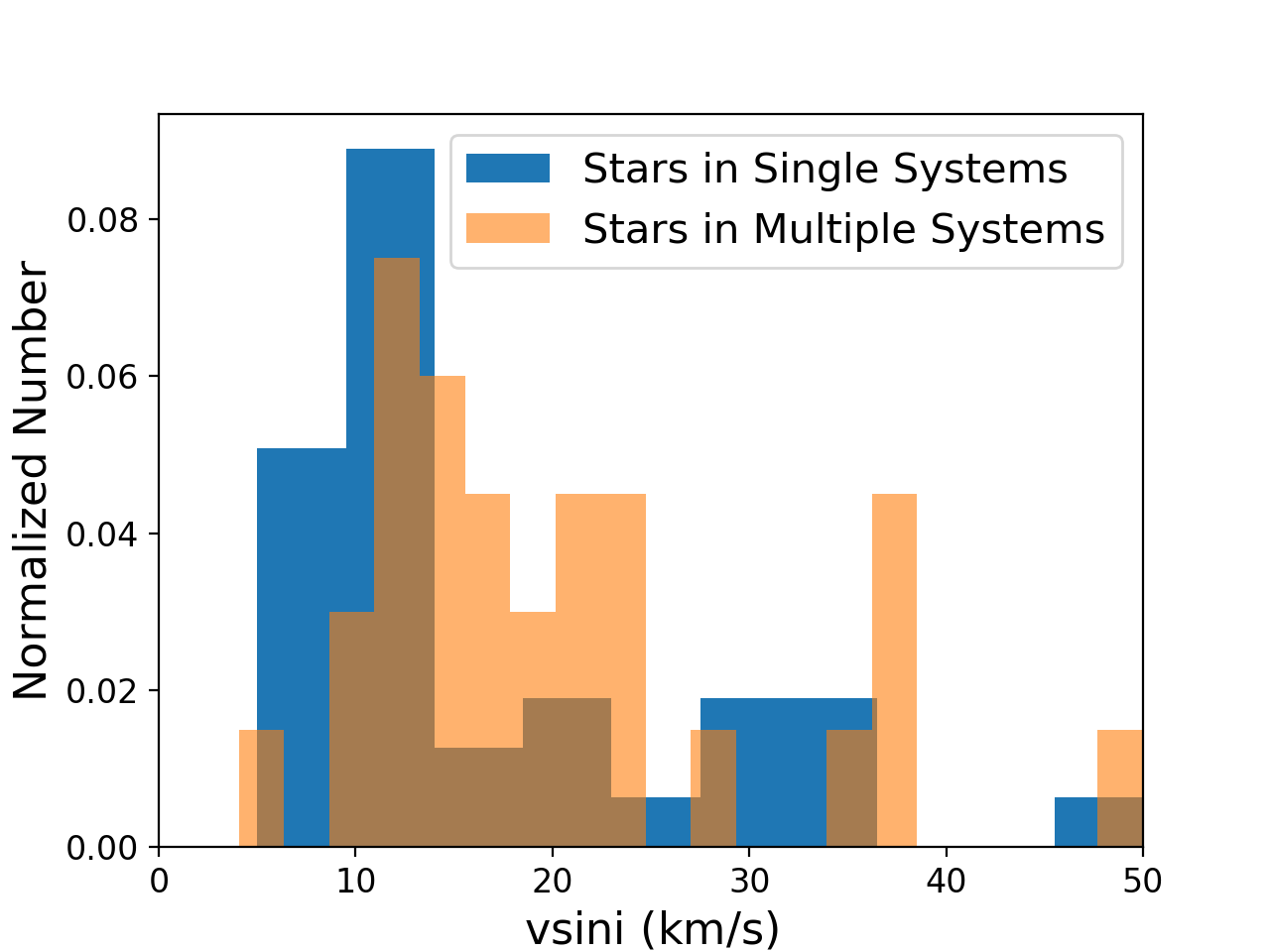}}
\caption{The $v \sin i$ distribution of single and multiple system pre-main sequence stars in our sample. Our $v \sin i$ measurements are higher on average for stars with stellar companions, which could be a consequence of contamination and line blending in the spectra, shorter disk lifetimes, or of tidal interactions in hierarchical triple systems. \label{fig:hist_sin_mul}}
\end{center}
\end{figure}

We used a Kolmogorov-Smirnov (K-S) and Anderson-Darling (A-D) test to confirm this result. The two-sample K-S test is a non-parametric method that compares two samples and determines whether they come from a single population. The k-sample A-D test is a modification of the K-S test and makes a similar comparison, however, the A-D test gives more weight to the tails of the distributions. These statistical tests determine a p-value, the probability of obtaining a result assuming that the distributions are the same. A K-S test and A-D test result in p-values of 0.056 and 0.054, respectively (Table \ref{tab:KS}). This result provides marginal support for rejecting the null hypothesis that assumes both samples come from the same population. The distributions are shown in Figure \ref{fig:hist_sin_mul}. 

We looked for evidence of contamination from a stellar companion for those targets that are part of a binary or multiple system and that have a higher $v \sin i$ than the sample average. First, we looked for obvious signs of companion contamination, such as a double-peak in the CCF. Another potential indicator of line blending from a stellar companion, particularly for equal-mass binaries, is temporal variation in the measured $v \sin i$ that could mimic rotational broadening (\citealt{1997ApJ...475..604S}). Spectra of long-period binaries may display wide lines that vary in width instead of distinct double lines. To check for this effect in our data, we investigated the variation in the measured $v \sin i$ values of the relevant targets. Three of the four observations of the CTTS binary, IT Tau ($v \sin i$ = 37$\pm$2 km s$^{-1}$), show a double-peak in the CCF indicating potential contamination from a stellar companion. IT Tau is a known double star (\citealt{1994ARA&A..32..465M}) with an angular separation of $\sim$ 2.5 mas and a K-band contrast ratio of 0.9 (\citealt{Kraus_2009}). We also note that the primary and companion should be barely resolved with our observing system, so it is possible there is contamination from the secondary. The WTTS binary LkH$\alpha$ 332/G1 ($v \sin i$ = 29$\pm$2 km s$^{-1}$) shows large differences in the $v \sin i$ measurements that could be temporal variation caused by a stellar companion. Based on separations measured by \cite{2012ApJ...745...19K}, we determine that this binary is not resolved with our observing system, therefore, contamination from the secondary could be the cause of the observed $v \sin i$ variation. Both IT Tau and LkH$\alpha$ 332/G1 were excluded from further analysis given the possibility that we did not measure the true $v \sin i$ due to secondary contamination. We also measured larger than average variations in the $v \sin i$ values for some CTTSs, regardless of multiplicity, which we attribute to veiling and low SNR of the spectral lines. However, these effects are not expected to impact the $v \sin i$ measurements of the WTTSs that are noted above. There was no direct evidence for contamination in the other multiple system spectra. 

Our above analysis suggests that the higher $v \sin i$ values for pre-main sequence binaries compared to single stars can likely not be entirely explained by measurement bias. Alternatively, it is possible that evolutionary factors may explain why stars in binary or multiple systems have a higher $v \sin i$ on average than those in single systems. Close binaries can be spun up by tidal interactions, leading to faster rotation rates (\citealt{1992ApJ...398L..61A}), and may experience truncation of the circumstellar disk, which would otherwise slow rotation (\citealt{2018AJ....156..275S}). Therefore, components of close binaries generally have rotation rates that are faster than those of single stars at young ages, however wider binaries are expected to have rotation rates similar to single stars (\citealt{2018AJ....156..275S}). Most of the companions in our sample have been confirmed using adaptive optics imaging (\citealt{2012ApJ...745...19K}), and are therefore too widely separated to experience tidal spin-up. It is possible that some of these binaries are actually hierarchical triples composed of a close inner binary pair and an outer companion. \citet{2006A&A...450..681T} determined that the fraction of spectroscopic binaries with an additional tertiary companion is $\sim$ 60-70$\%$, making these types of systems relatively common. We conclude that evolutionary factors involving a companion close enough to influence the primary circumstellar disk may be responsible for the higher than average $v \sin i$ of some of our targets. 

\subsection{Comparison of Optical and Infrared $v \sin i$}

\subsubsection{Comparison to Optical Values}

\begin{figure}
\begin{center}
\resizebox{\hsize}{!}{\includegraphics{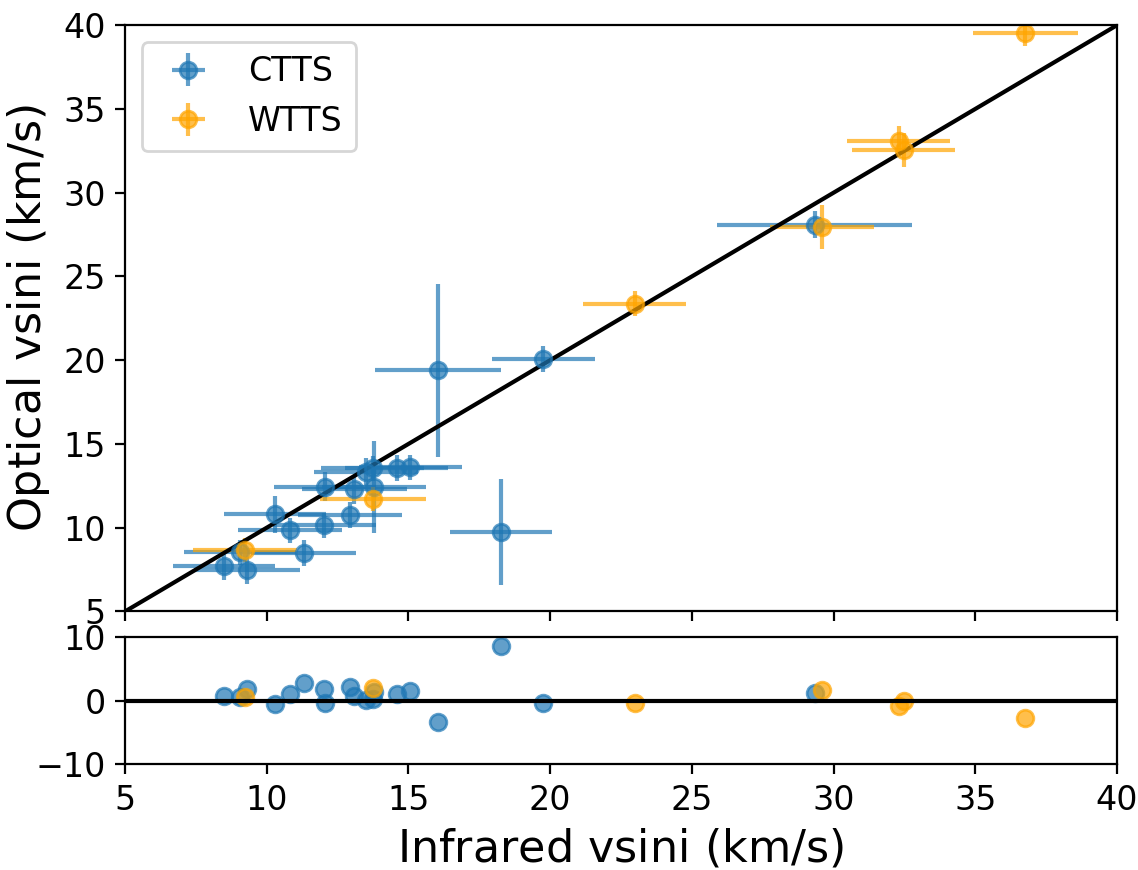}}
\caption{Comparison of the optical and infrared $v \sin i$ measurements and their residuals. The most significant outlier (XZ Tau, bottom center) is a heavily veiled target. A K-S test and an A-D test both indicate that there is no significant difference between the infrared and optical $v \sin i$ distributions. \label{fig:ir_opt}}
\end{center}
\end{figure}

We made a direct comparison between $v \sin i$ values measured at optical and infrared regimes. In addition to the difference in wavelengths, these $v \sin i$ values were calculated from observations taken during different epochs and using different instruments. The analysis techniques also differed; the optical measurements were made using observed stellar templates and the LSD technique, and the infrared measurements relied on synthetic stellar models (see Sections \ref{sec:analysis_ir} and \ref{sec:analysis_op} for analysis details). A direct comparison of the results from two different methods at two different wavelengths can confirm our error estimates or identify astrophysical reasons for discrepancies between the two datasets. Figure \ref{fig:ir_opt} shows a comparison between the optical and infrared $v \sin i$ values and the residuals. The $v \sin i$ values above the estimated upper limit for fast rotators were excluded from the comparison. In general, there is good agreement between these two datasets. The most significant outlier (bottom center) is XZ Tau, which is known to be heavily veiled (\citealt{2001ApJ...556..265W}). While veiling does not broaden or narrow the FWHM of the CCF, the continuum emission caused by warm dust weakens the stellar absorption lines resulting in a CCF that lacks clear structure. This can lead to an inaccurate $v \sin i$ calculation. Published results from \citet{2012ApJ...745..119N} determined a $v \sin i$ value of $\sim$ 15 km s$^{-1}$ for XZ Tau, which falls in between the infrared value of $\sim$ 18 km s$^{-1}$ and the optical value of $\sim$ 10 km s$^{-1}$.

We statistically compared the 26 infrared and optical $v \sin i$ values within the established measurement limits. A K-S test and an A-D test indicates a p-value of 0.440 and 0.405, respectively (Table \ref{tab:KS}), which shows that the null hypothesis is not rejected. Thus there is no significant difference between the optical and infrared $v \sin i$ distributions.

This result allows us to place limits on potential causes of variations between the infrared and optical measurements. The observed wavelength regimes may be responsible for some of the variation. Infrared observations could be affected by contamination of the CO lines from nonstellar sources, such as line emission caused by high velocity disk material and warm dust in the circumstellar disk (\citealt{1983ApJ...275..201S, 1985ApJ...299L..41T, 1989ApJ...345..522C, 1991ApJ...380..617C, 2001ApJ...561.1060J}). Depending on the mass accretion and irradiation rates of the system, the effect of the CO contribution from the disk either decreases the strength of the absorption lines in the CO bands or converts them into emission (\citealt{1991ApJ...380..617C}). For the vast majority of CTTSs, however, the CO lines in the K-band were found to originate in the stellar photosphere rather than the disk (\citealt{1996A&A...306..427C, 2001ApJ...561.1060J, 2007ApJ...664..975J}).

To test whether there is a significant disk CO contribution in the CTTS systems, we also separately compared infrared and optical $v \sin i$ values for CTTSs and WTTSs (Figure \ref{fig:ir_opt}). If there is significant CO emission from the disks in the CTTS systems, then we expect to find a larger difference between the measurements at the two wavelengths for the CTTS subset. A K-S test of the CTTS infrared and optical $v \sin i$ values shows that they are not statistically distinguishable, with a p-value of 0.526. Similarly, the WTTS subset has a p-value of 0.938. An A-D test indicates a p-value for the CTTS sub-sample of 0.336 and 0.990 for the WTTS sub-sample. Table \ref{tab:KS} summarizes the K-S and A-D test results between the infrared and optical measurements. Based on this result, we do not see significant evidence for CO contamination in the CTTS sub-sample. 

\subsubsection{Comparison to Literature Values \label{sec:lit}}

\begin{figure}
\begin{center}
\resizebox{\hsize}{!}{\includegraphics{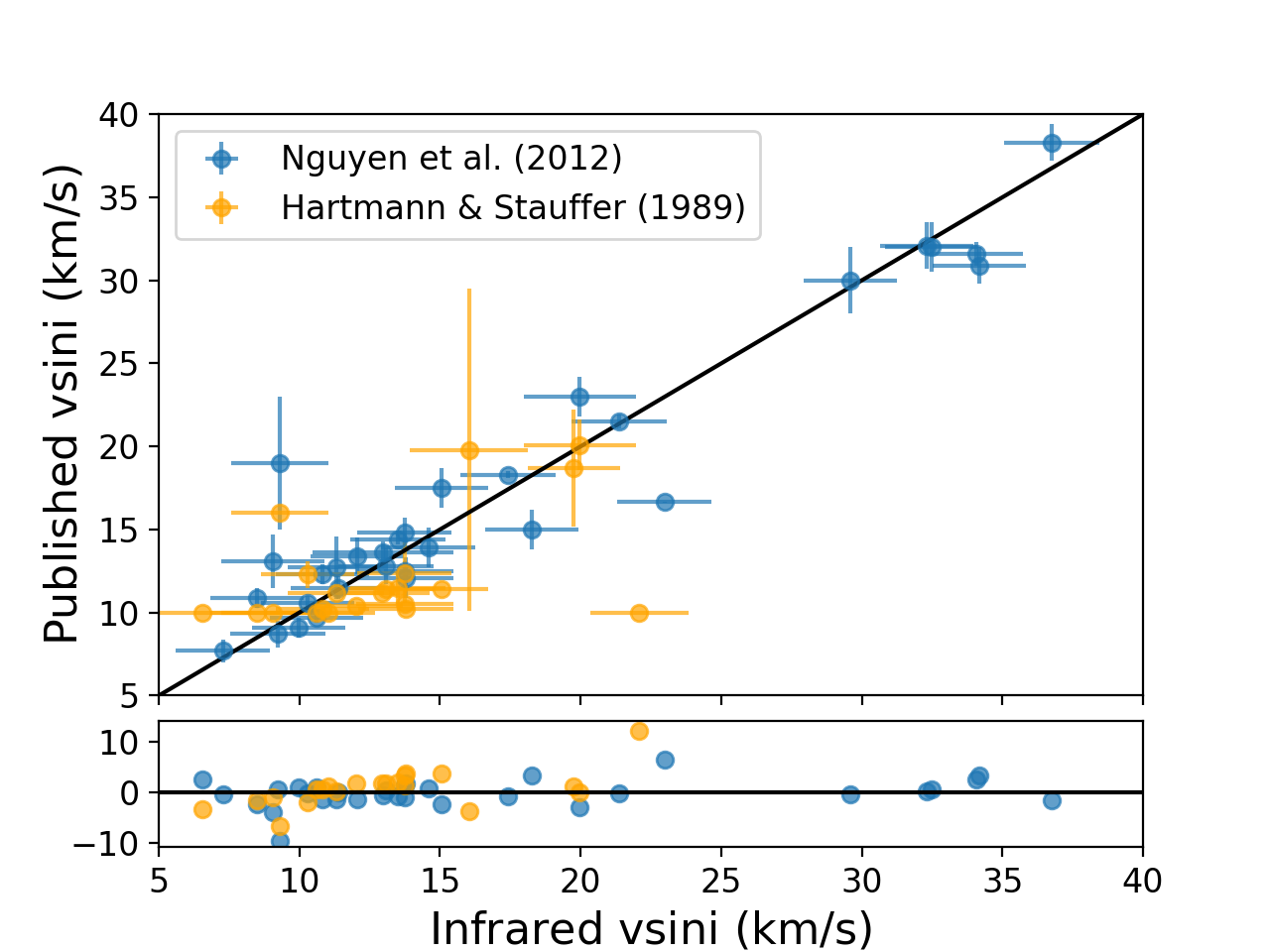}}
\caption{Comparison of the infrared $v \sin i$ measurements with published optical measurements from the literature. A K-S and A-D test both indicate no statistically significant difference between the infrared and optical $v \sin i$ measurements. \label{fig:compare_lit}}
\end{center}
\end{figure}

To confirm the result in the previous section we also made comparisons between the infrared $v \sin i$ measurements and literature values determined from optical observations from \citet{1989AJ.....97..873H} and \citet{2012ApJ...745..119N}. Figure \ref{fig:compare_lit} shows a comparison to the $v \sin i$ values in \citet{1989AJ.....97..873H} for an intersecting sub-sample of 21 targets, and excluding targets with rotation outside our limits. We observe agreement within $\sim$ 1-3 km s$^{-1}$ for $\sim$ 85$\%$ of the sub-sample. For the remaining $\sim$ 15$\%$, we have identified these targets as either being heavily veiled (DL Tau), residing in a multiple system (HK Tau), or having a $v \sin i$ near the resolution limits of the two surveys (DM Tau). Many of the $v \sin i$ estimates from \citet{1989AJ.....97..873H} do not have uncertainties, and a K-S test and A-D test do not show a significant difference between the two samples with p-values of 0.135 and 0.207, respectively.

We further compared our $v \sin i$ measurements to those from \citet{2012ApJ...745..119N} for the intersecting subset of 32 targets. Table \ref{tab:properties} lists the \citet{2012ApJ...745..119N} $v \sin i$ values. K-S and A-D tests indicate no statistical difference between the two sub-samples of $v \sin i$ measurements, with p-values of 0.795 and 0.812, respectively (Table \ref{tab:KS}). This result shows no significant difference between $v \sin i$ derived from analysis using template or synthetic spectra, and thus supports the use of either method to accurately measure rotational velocities. Our optical measurements are also mostly consistent with the \citet{2012ApJ...745..119N} values for an intersecting subset of 22 targets: 55$\%$ of our sub-sample matches to within 1-$\sigma$, while 91$\%$ matches to within 3-$\sigma$. A K-S and A-D test result in p-values of 0.175 and 0.086, respectively, and therefore also do not show a significant difference between the samples. 

\begin{figure}
\begin{center}
\resizebox{\hsize}{!}{\includegraphics{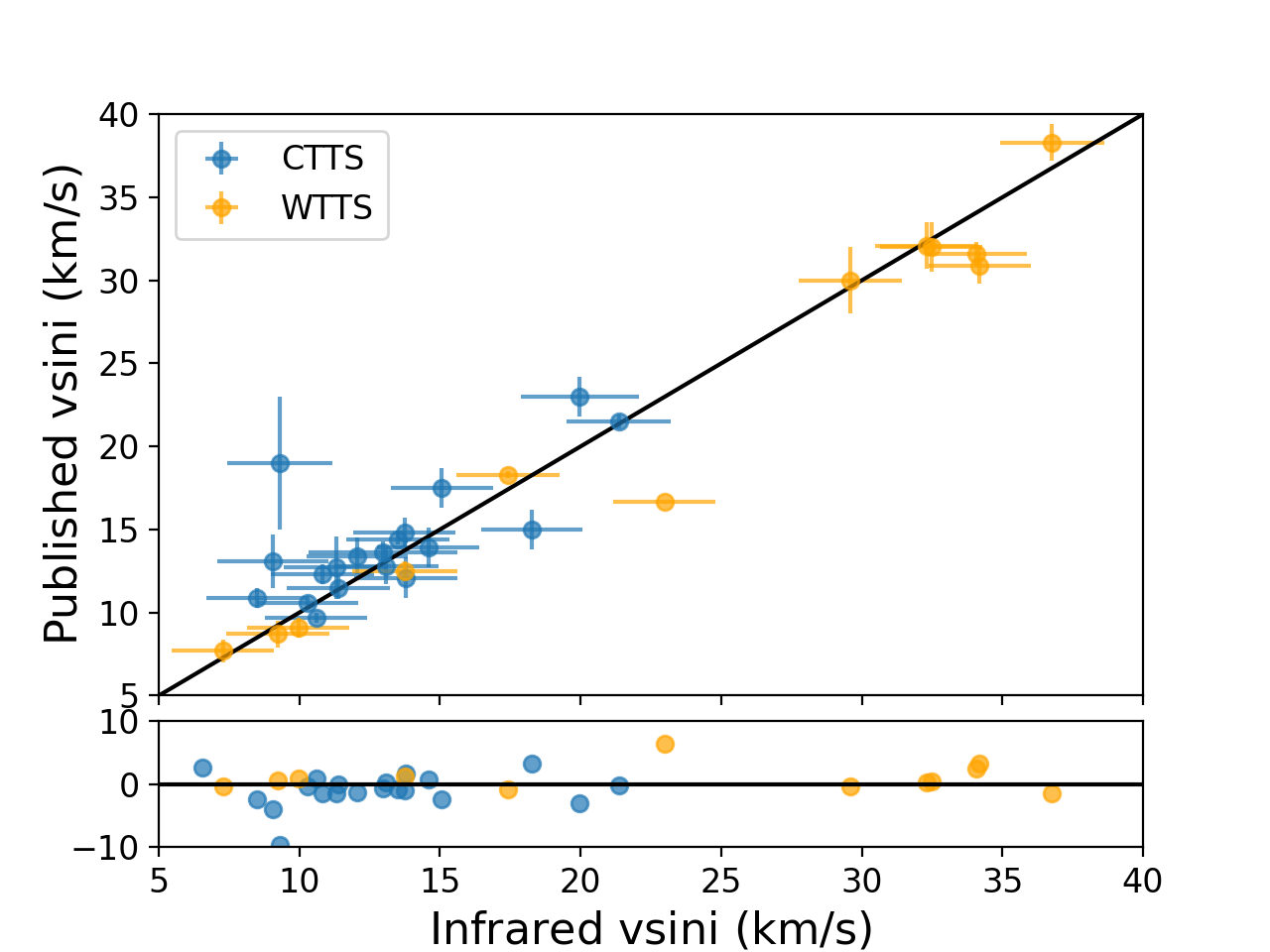}}
\caption{Comparison of the infrared $v \sin i$ measurements with published values from \citet{2012ApJ...745..119N}. A K-S and A-D test indicate no statistical difference between the infrared and published $v \sin i$ measurements for the CTTS sample, or for the WTTS sample. \label{fig:compare_lit_class}}
\end{center}
\end{figure}

To further investigate whether the infrared $v \sin i$ measurements are significantly affected by a contribution of CO from the disk, we compared the infrared and \citet{2012ApJ...745..119N} values separately for the 20 CTTSs and 12 WTTSs in the overlapping sample (Figure \ref{fig:compare_lit_class}). To confirm whether there is indeed a better match for WTTSs, we repeated the statistical tests on both the CTTS and WTTS sub-samples. A K-S and A-D test of the CTTS infrared and optical $v \sin i$ values shows that the samples are not statistically distinguishable with a p-value of 0.497 and 0.250, respectively (Table \ref{tab:KS}). This result is consistent with the K-S test result for the infrared and optical CTTS $v \sin i$ comparison. For the WTTS subset, the distributions show no statistical difference, with a p-value of 0.433 (K-S test) and 0.250 (A-D test) (Table \ref{tab:KS}). Therefore, a comparison between the infrared $v \sin i$ values and published optical $v \sin i$ values, shows no evidence that CO contamination from the disks significantly affects the infrared $v \sin i$ measurements. 

\subsection{Correlation of $v \sin i$ and Evolutionary States \label{sec:evolution}}

\subsubsection{CTTS and WTTS Distributions}

Previous studies have investigated whether there is a significant difference between the $v \sin i$ distributions of CTTSs and WTTSs (\citealt{1988AJ.....96..297W}; \citealt{1989AJ.....97..873H}; \citealt{1993A&AS..101..485B}; \citealt{1993AJ....106..372E}; \citealt{1996AJ....111..283C}; \citealt{2001AJ....122.3258R}; \citealt{2002A&A...396..513H}; \citealt{2005AJ....129..363S}; \citealt{2009ApJ...695.1648N}). CTTSs are prevented from spinning up as they contract due in part to magnetic disk-braking, but WTTSs no longer have a disk (or only have a tenuous disk) and spin-up as they contract. Therefore, stellar evolution theory predicts that stars with disks rotate more slowly than those without (\citealt{2012ApJ...745...56D}), resulting in three scenarios: slow rotators with disks, slow rotators without disks that have not yet spun-up, and fast rotators without disks. For the most part, we do not expect to identify any fast rotators with disks (\citealt{2009ApJ...695.1648N}), although there can be exceptions, such as young sources that have not yet established disk-locking (\citealt{2005A&A...430.1005L}; \citealt{2012ApJ...756...68C}), stars that have been spun-up from a brown dwarf merger (\citealt{2002MNRAS.330L..11A}), or exceptions based on stellar magnetic field strength and disk accretion rates (\citealt{2002ApJ...573..685J}). Many previous studies have found a statistically significant difference between the $v \sin i$ distribution of CTTSs and WTTSs, with WTTSs rotating faster than CTTSs (\citealt{1988AJ.....96..297W}; \citealt{1993A&AS..101..485B}; \citealt{1993AJ....106..372E}; \citealt{1996AJ....111..283C}; \citealt{2002A&A...396..513H}; \citealt{2005AJ....129..363S}), yet other surveys have found that these distributions are actually statistically the same (\citealt{1989AJ.....97..873H}; \citealt{2001AJ....122.3258R}) or that the results are inconclusive (\citealt{2009ApJ...695.1648N}). 

\begin{figure}
\begin{center}
\resizebox{\hsize}{!}{\includegraphics{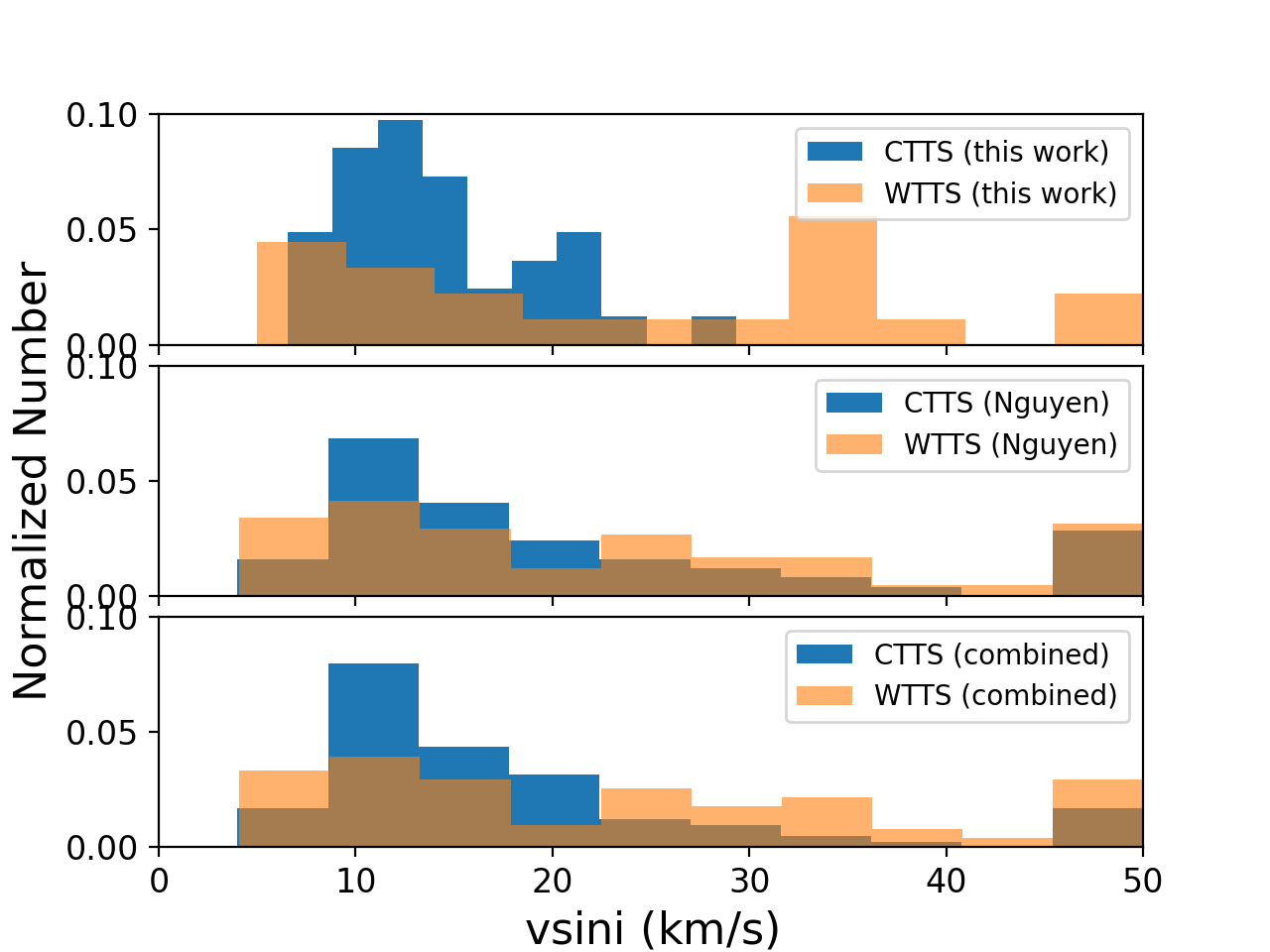}}
\caption{The $v \sin i$ distributions for CTTSs and WTTSs in our sample (top panel), a sample chosen from \citet{2012ApJ...745..119N} (middle panel), and the combination of both samples (bottom panel). All $v \sin i$ values above the upper limit of 50 km s$^{-1}$ are plotted at this limit. A K-S and A-D test show that these distributions are statistically different in our sample. The sample from the literature does not show evidence of this difference, as indicated by statistical tests. The combined sample shows that the $v \sin i$ distributions for CTTSs and WTTSs are statistically distinguishable. Therefore, this difference in our dataset is likely caused by a selection bias. \label{fig:hist_class}}
\end{center}
\end{figure}

A histogram comparing the CTTS and WTTS $v \sin i$ distributions of our sample is shown in the top panel of Figure \ref{fig:hist_class}. The distributions suggest that there are populations of slow-rotating stars with disks, slow-rotating stars without disks, and fast-rotating stars without disks, as expected by stellar evolution theory. There are a few exceptions in the CTTS distribution that have $v \sin i$ $>$ 20 km s$^{-1}$, which may be due to the reasons noted above. The WTTSs in this sample are generally faster rotators than the CTTSs. The four targets that have been identified as belonging to an older population (Section \ref{sec:sample}; \citealt{2018AJ....156..271L}) are included in this analysis: all are WTTSs with $v \sin i$ of 6, 16, 36, and $>$ 50 km s$^{-1}$, respectively. Since the $v \sin i$ estimates span a wide range of values, and this analysis compares the $v \sin i$ distributions of pre-main sequence stars based on classification rather than age, they were not omitted from the comparison. A K-S test shows that the distributions are statistically different with a p-value of 0.002; an A-D test confirms this result with a p-value of 0.003 (Table \ref{tab:KS}). While this outcome supports the expectation of $v \sin i$ evolution of the pre-main sequence stars in our sample, we caution that our sample selection may also explain this result. The sample was primarily chosen to include slow rotators ($v \sin i$ $<$ 20 km s$^{-1}$), and additional targets were chosen from a sample predominantly comprised of WTTSs. Therefore, the target selection ensures that the fast rotators in the sample are more likely to be WTTSs rather than a randomly chosen subset of CTTSs and WTTSs. 

As shown in Section \ref{sec:lit}, a comparison between our measured $v \sin i$ values and those from \citet{2012ApJ...745..119N} were consistent for a subset of overlapping targets. Therefore, to further explore this result, we increased our sample size by including 52 additional CTTSs and 78 WTTSs from \citet{2012ApJ...745..119N}. Only the targets with $v \sin i$ $<$ 50 km s$^{-1}$ were included since this is the upper limit on the $v \sin i$ that we can accurately measure and confirm. The middle panel of Figure \ref{fig:hist_class} shows the \citet{2012ApJ...745..119N} $v \sin i$ distributions for CTTSs and WTTSs. \citet{2009ApJ...695.1648N} showed that there is no statistically significant difference between the CTTS and WTTS vsini distributions in their sample. We repeated this statistical analysis on the \citet{2012ApJ...745..119N} sample, which includes an overlapping but larger number of targets than that presented in \citet{2009ApJ...695.1648N}, and found the same result. The chosen subset from \citet{2012ApJ...745..119N} does not show that the CTTS and WTTS $v \sin i$ distributions are different, with a K-S test and A-D test indicating p-values of 0.177 and 0.179, respectively (Table \ref{tab:KS}). However, combining the subset of additional $v \sin i$ measurements with our values (bottom panel of Figure \ref{fig:hist_class}) results in two distributions that are statistically distinguishable, with a p-value of 0.028 (K-S test) and 0.015 (A-D test) (Table \ref{tab:KS}).

While the combined sample also demonstrates a statistical difference between the CTTS and WTTS $v \sin i$ distributions, our selection bias may still be the cause of this result. Both samples are magnitude-limited, with V $<$ 15 and R $\leq$ 13.4 for our sample and the \citet{2012ApJ...745..119N} sample, respectively. The \citet{2012ApJ...745..119N} sample, however, differs from ours in that they excluded all unresolved or spectroscopic binaries, which limits artificially high $v \sin i$ caused by contamination. They also included targets with a larger range of masses, of spectral types F2-M5, while our sample is limited to K and M stars. The inclusion of higher mass targets is expected to increase the $v \sin i$ in the distribution because $v \sin i$ is correlated with spectral type, whereas stars more massive than the Sun typically rotate more rapidly than low-mass stars (\citealt{1981ApJ...245..960V, 1989AJ.....97..873H, 2009ApJ...695.1648N, 2012ApJ...745...56D}). Combining these samples does not conclusively determine if the statistical difference is an evolutionary effect, particularly when considering a wider range of masses. The conflicting results between the two samples reinforces the conclusion that the statistically significant difference in the CTTS and WTTS $v \sin i$ distributions that we detected is likely due to a selection bias.  

\subsubsection{Evolution on the H-R Diagram}

The $v \sin i$ of a pre-main sequence star is expected to change as the star evolves towards the main sequence, first by decreasing due to interactions with the circumstellar disk, then by increasing after the disk disperses and the star continues to contract. 

\begin{figure*}
\epsscale{1.17}
\plottwo{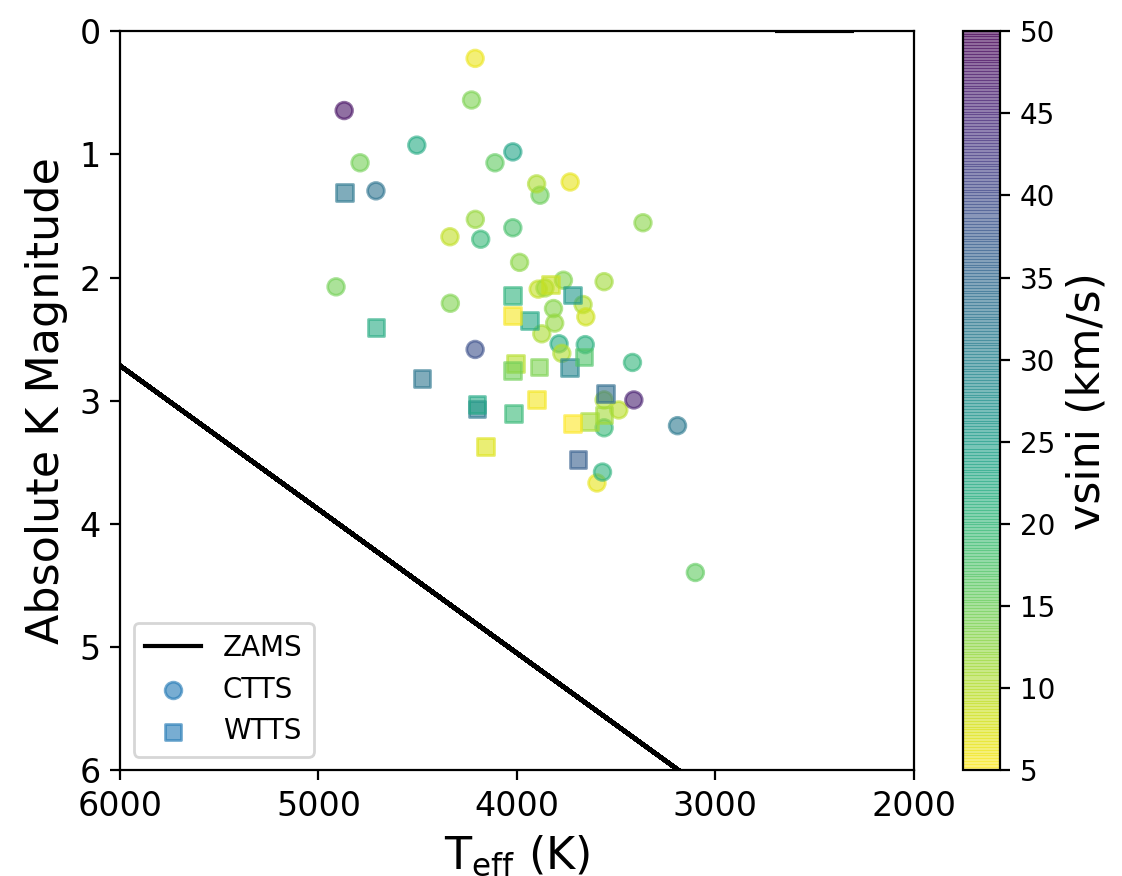}{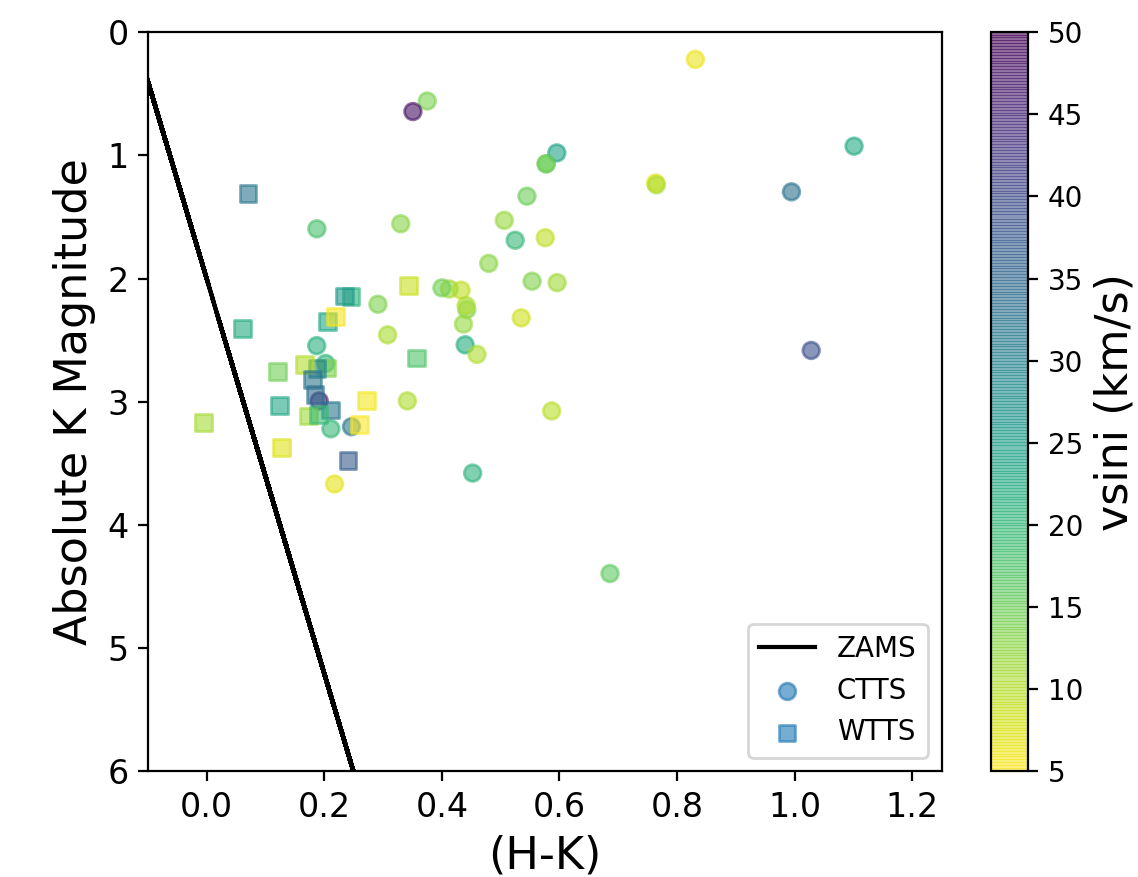}
\caption{A Hertzsprung-Russell diagram (left) and a color-magnitude diagram (right) of the 65 pre-main sequence stars in the combined sample that are confirmed Taurus members and have measurements of H and K magnitudes, extinction corrections, parallaxes, and $T_{\rm eff}$. This sample includes CTTSs and WTTSs with the $v \sin i$ indicated by color. A line indicating the estimated position of stars from the Pleiades open cluster is overplotted to demonstrate the evolutionary stage of the pre-main sequence star sample relative to the zero-age main sequence. Typical error bars are shown in the upper left (the absolute K-magnitude error is smaller than the symbol sizes). There are no obvious trends based on the stellar classification or $v \sin i$. \label{fig:hrd}}
\end{figure*}

To investigate the correlation between the $v \sin i$ and stellar evolution, we plotted our sample on a Hertzsprung-Russell diagram (H-R diagram) and a near-IR color-magnitude diagram (CMD) in Figure \ref{fig:hrd}. We collected H and K magnitudes, extinction corrections, and parallaxes (from which distances and absolute magnitudes could be calculated), from the database presented in \citet{2019AJ....158...54E}. This database includes H and K magnitudes from the United Kingdom Infrared Telescope (UKIRT) Infrared Deep Sky Survey (UKIDSS; \citealt{2007MNRAS.379.1599L}) and \textit{Gaia} DR2 parallaxes, as well as J band extinction estimates (A$_{\rm J}$) and membership information. All H and K magnitudes were corrected for extinction using the A$_{\rm J}$ estimates in combination with the A$_{\rm J}$/A$_{\rm H}$ and A$_{\rm J}$/A$_{\rm K}$ ratios from \citet{2005ApJ...619..931I}. We included 42 targets from our sample and combined them with our $T_{\rm eff}$ measurements. We supplemented this sample with 23 additional targets from \citet{2012ApJ...745..119N}. All targets included in the H-R diagram and CMD analysis met the following criteria: confirmed membership in Taurus, measured H and K magnitudes, extinction estimates, $T_{\rm eff}$ estimates (either from our analysis or the literature), and \textit{Gaia} parallaxes. The parallaxes allowed us to obtain distances to calculate absolute magnitudes. Since the $v \sin i$ was shown to be consistent between our values and the \citet{2012ApJ...745..119N} values for an overlapping subset of targets (Section \ref{sec:lit}), we included all additional targets meeting the above criteria, and  with measured $v \sin i$ $<$ 50 km s$^{-1}$, which is the limit at which we no longer accurately measure $v \sin i$. The $T_{\rm eff}$ of the additional targets was estimated from the spectral types presented in \citet{2012ApJ...745..119N}, and the \citet{2014ApJ...786...97H} spectral type to temperature conversions. Overplotted on both diagrams is an estimate of the zero-age main sequence based on a subset of Pleiades targets ($\sim$ 100 Myr old; \citealt{1993A&AS...98..477M}), for reference. The Pleiades sequence was collapsed using quantile regression, which estimates the median sequence instead of the mean. The Pleiades H and K magnitudes were measured by \cite{2015yCat..35770148B} and the \textit{Gaia} DR2 distance estimates were determined by \cite{Abramson_2018}. 

Figure \ref{fig:hrd} shows no obvious relation between H-R diagram position, $v \sin i$ and the presence of accretion signatures. To quantify potential trends between stellar properties and position on the H-R diagram and CMD, we calculated the distance of each object on the diagrams relative to the zero-age main sequence and compared the average and median distances based on classification (CTTS/WTTS) and rotation (slow rotator with $v \sin i$ $<$ 20 km s$^{-1}$ and fast rotator with $v \sin i$ $>$ 20 km s$^{-1}$). The WTTSs have an average distance to the zero-age main sequence that is $\sim$ 10$\%$ closer than the CTTS average distance in the H-R diagram, and $\sim$ 20$\%$ closer in the CMD. However, the estimated uncertainties on these measurements are 10-15$\%$, and the robust median values indicate that WTTSs are only closer than the CTTSs to the zero-age main sequence by 1-2$\%$. Slow rotators are negligibly closer to the zero-age main sequence than fast rotators in the H-R diagram (by $\sim$ 1$\%$), but slow rotators are $\sim$ 30$\%$ closer on average in the CMD ($\sim$ 20$\%$ considering the median values). These differences are still within the uncertainties of the distance measurements, and therefore not statistically significant. The apparent evolution of low-mass stars to the zero-age main sequence does not proceed at the same rate for all pre-main sequence stars, and is dependent on local initial conditions (\citealt{1988AJ.....96..297W}). Our results support this conclusion and indicate a complex interplay between stellar rotation and the evolution of pre-main sequence stars towards the zero-age main sequence. 

It is important to note that inferring evolutionary states from the placement of pre-main sequence stars on the H-R diagram or CMD is limited. Magnitudes are affected by photometric variability caused by accretion (e.g., \citealt{2017MNRAS.468..931M}), stellar luminosity and $T_{\rm eff}$ are affected by starspots (e.g., \citealt{2015ApJ...807..174S, 2019ApJ...882...75F}), and Li abundance as an age indicator, as well as other astrophysical measurements, are complicated by radius inflation (\citealt{2014ApJ...790...72S}). These phenomena all have an effect on the placement of pre-main sequence stars on evolutionary tracks. 

\section{Conclusions} \label{sec:conclusions}

We have presented stellar properties for 70 low-mass pre-main sequence stars in the Taurus-Auriga region. These stellar properties include projected rotational velocities, effective temperatures, rotation periods, and limits on radii. About 70$\%$ of our sample show evidence of having an accreting circumstellar disk, and just under half the sample are known multiples. We calculated $T_{\rm eff}$ using infrared spectra and $T_{\rm eff}$-LDR relations (\citealt{2019ApJ...879..105L}), $v \sin i$ using cross-correlation analysis of infrared and optical spectra, P$_{\rm rot}$ from ground-based and space-based photometry and published values, and $R \sin i$ from a relation between $v \sin i$ and P$_{\rm rot}$. Our main scientific conclusions are as follows:

\begin{itemize}
  \item Infrared $v \sin i$ measurements of pre-main sequence stars calculated using the cross correlation technique agree with those measured by an independent MCMC technique, with a typical scatter of $\sim$ 1.8 km s$^{-1}$. This defines a typical error floor for the $v \sin i$ of these stars from infrared spectra.
  \item A comparison of the $v \sin i$ distributions of stars with and without companions shows a significant difference, with binaries/multiples typically having higher measured $v \sin i$. This can partly be explained by line broadening from companion spectral lines. Tidal interactions in hierarchical triples may also affect some of these systems.
  \item A comparison of optical and infrared $v \sin i$ values shows no statistical difference regardless of whether the star has a disk or not. This indicates that CO contamination from the disk does not significantly impact $v \sin i$ measurements above the typical error floor of our measurements ($\sim$ 1.8 km s$^{-1}$). 
  \item We do not see any clear correlation between $v \sin i$, presence of an accreting disk, and H-R diagram position, which indicates a complex interplay between stellar rotation and evolution of pre-main sequence stars. Based on average distance measurements, we found that WTTSs were $\sim$ 10$\%$ and $\sim$ 20$\%$ closer than the CTTSs to the zero-age main sequence in the H-R diagram and CMD, respectively, while slow rotators ($v \sin i$ $<$ 20 km s$^{-1}$) were about 30$\%$ closer than rapid rotators to the zero-age main sequence in the CMD. However, these results are within the uncertainties of the distance measurements and do not conclusively determine any trends between stellar rotation or classification and evolutionary state. 
\end{itemize}

The results presented here illustrate the importance of high-resolution infrared spectroscopy to characterize pre-main sequence stars. Future work will include a larger survey of YSO properties (L\'{o}pez--Valdivia et al. in prep.). While large uncertainties ($\sim$ 200 K) in $T_{\rm eff}$ do not significantly affect our results, the effects of the magnetic field and veiling should be considered in the determination of the stellar parameters ($T_{\rm eff}$, $\log g$, $v \sin i$) to try to  break possible degeneracies. The stellar properties presented in this work were determined as part of an infrared radial velocity survey to find and characterize young hot Jupiters and other substellar companions.

\acknowledgments

L.N. thanks Karen Meech, Philip Massey, and Nicholas Moskovitz for valuable discussions. D.H. acknowledges support from the Alfred P. Sloan Foundation. J.T. acknowledges support for this work was provided by NASA through the NASA Hubble Fellowship grant $\#$51424 awarded by the Space Telescope Science Institute, which is operated by the Association of Universities for Research in Astronomy, Inc., for NASA, under contract NAS5-26555. We acknowledge the efforts of the Lowell Discovery Telescope and McDonald Observatory staff, and observers Nolan Habel and Nicole Karnath. Partial funding for this project is provided by NASA XRP award 8NSSC19K0289 (PI Prato). L.N. would like to acknowledge the gracious support of her Lowell Pre-doctoral Fellowship by the BF Foundation, the Visiting Astronomer at the Infrared Telescope Facility program, which is operated by the University of Hawaii under contract 80HGTR19D0030 
with the National Aeronautics and Space Administration, and the Institute for Astronomy and University of Hawaii for supporting a graduate scholarship. This work used the Immersion Grating Infrared Spectrometer (IGRINS) that was developed under a collaboration between the University of Texas at Austin and the Korea Astronomy and Space Science Institute (KASI) with the financial support of the US National Science Foundation under grants AST-1229522 and AST-1702267, of the University of Texas at Austin, and of the Korean GMT Project of KASI. This paper includes data taken at the McDonald Observatory of The University of Texas at Austin. These results made use of the Lowell Discovery Telescope at Lowell Observatory. Lowell is a private, non-profit institution dedicated to astrophysical research and public appreciation of astronomy and operates the LDT in partnership with Boston University, the University of Maryland, the University of Toledo, Northern Arizona University and Yale University. 

\bibliography{vsini.bib}

\end{document}